\documentclass[journal]{IEEEtran}
\usepackage{amsmath}
\usepackage{amssymb}
\usepackage{mathtools}
\usepackage{bm}
\usepackage{ulem}
\usepackage{amsthm}
\usepackage{algorithm}
\usepackage{algpseudocode}
\usepackage{graphicx}
\usepackage{subcaption}
\usepackage{float}
\usepackage{booktabs}
\usepackage{pifont}   
\newcommand{\cmark}{\ding{51}}
\newcommand{\pmark}{\scriptsize (partial)}
\usepackage{array}
\usepackage{newtxtext}
\usepackage{newtxmath}
\usepackage[dvipsnames]{xcolor}
\usepackage[colorlinks=true,
            linkcolor=blue,
            citecolor=blue,
            urlcolor=blue]{hyperref}
\usepackage{enumitem}
\usepackage{url}
\usepackage{cite}

\begin{document}

\title{Security Engineering of OpenClaw: Analyzing Attack Surface Expansion and Trust-Boundary Violations}

\author{
Saeid~Jamshidi,
Arghavan~Moradi~Dakhel,
Kawser~Wazed~Nafi and
Foutse~Khomh
\thanks{
S. Jamshidi, A. Moradi Dakhel, K. W. Nafi, and F. Khomh are with the SWAT Laboratory,
Polytechnique Montréal, Montréal, QC, Canada
(e-mail: \{saeid.jamshidi, arghavan.moradi-dakhel, kawser.wazed-nafi, foutse.khomh\}@polymtl.ca).
}%
}

\maketitle

\begin{abstract}
Agentic large language model (LLM) systems can now execute actions, not just produce text. When model outputs directly trigger privileged operations such as shell commands, browser automation, and external tool invocation, the security problem shifts from alignment concerns to system configuration and structural design. We analyze OpenClaw, a self-hosted multi-agent system in which LLM outputs can execute commands and interact with system tools and services. We measure compromise probability, boundary failures, privilege drift, and how these metrics change as the attacker's capability increases. With one agent, the probability of compromise is 0.24. When seven agents are active, and the system executes an action if any single agent proposes it, compromise increases to 0.86. The models themselves do not change; the increase stems from how their outputs are aggregated. Prompt injection propagates instability across the system. Attack surface entropy increases from 0.42 to 0.71, reflecting a wider distribution of potential exploit paths. Mean privilege drift rises from 0.03 to 0.21, indicating more substantial unintended authority gain. The escalation curvature is positive (0.08), indicating that privilege grows faster as the attacker's capability increases. Defensive controls, including policy gating and execution filtering, reduce compromise probability by 0.10, boundary failures by 0.10, and privilege drift by 0.02 (all statistically significant, $p < 0.0001$). The system remains sensitive, yet the mitigation impact is measurable. Injection mitigation success differs across models: 0.37 for GPT-5.2, 0.35 for Llama-4-Maverick, and 0.31 for DeepSeek-R1. When execution can be triggered by any single agent, the agent with the highest vulnerability determines the system's exposure. Mitigation strategies slightly reduce task utility (from 0.93 to 0.89) and increase median latency (from 420 ms to 468 ms). 
\end{abstract}


\section{Introduction}
\label{sec:introduction}
Large Language Models (LLMs) are rapidly advancing beyond simple text generation, evolving toward autonomous decision-making and task management capabilities through the emergence of agentic AI frameworks \cite{brohi2025research} \cite{jiang2026large} \cite{alva2026agentic}. In agentic settings, models interpret context, select tools, and execute actions that directly impact external systems \cite{miao2025towards}\cite{annepaka2025large}\cite{rane2023contribution}. APIs, file systems, network services, and third-party platforms are now integrated into the impact loop, allowing model outputs to trigger privileged operations \cite{luo2025large} \cite{plaat2025agentic}\cite{barua2024exploring}. In this setting, the security problem extends beyond alignment failures in text generation. Reasoning becomes coupled to execution, and model outputs act as operational decisions rather than terminal responses \cite{liu2023trustworthy}\cite{xu2024large}\cite{habibzadeh2025large}. Prior work has established that prompt injection and indirect manipulation remain persistent weaknesses in tool-enabled agents. InjecAgent demonstrates that untrusted contextual inputs can impact tool selection across heterogeneous environments \cite{zhan2024injecagent}. Recent studies evaluate attack success, defensive robustness, and benign task performance in realistic agent workflows \cite{debenedetti2024agentdojo}, confirming that injection attacks can remain impactful even when models appear aligned under static evaluation. Other research explores architectural countermeasures that separate reasoning from actuation to reduce the impact of injection \cite{debenedetti2025defeating}, while survey studies on multi-agent LLM systems highlight the complexity of coordination and the risks of shared execution when multiple models operate on common substrates \cite{wang2024survey}. These survey studies show that vulnerabilities in agentic systems are shaped not only by models themselves but also by system structure and execution design. What remains underexamined is how the configuration of multi-agent systems reshapes exposure when reasoning is directly connected to execution, including the impact of proposal aggregation, trust boundaries, and policy enforcement on system-level risk. This paper addresses that gap through a structural security analysis of OpenClaw, a self-hosted multi-agent LLM system in which semantic outputs are mapped directly to executable primitives. Rather than treating prompt injection as an isolated vulnerability, we examine how the combination of multiple agents, boundary controls, and escalating attacker capabilities shapes the probability of compromise and the dynamics of privilege. We introduce quantitative metrics for system-level compromise amplification, cumulative boundary instability (CBI), privilege drift, attack surface entropy, and escalation curvature. Our empirical results demonstrate clear structural shifts. With a single agent, the probability of compromise is 0.24. When seven agents are active and execution proceeds if any one agent proposes an action, compromise probability increases to 0.86, corresponding to a 3.58$\times$ amplification. Injection raises attack-surface entropy from 0.42 to 0.71, indicating a diffusion of risk across internal boundaries. Mean privilege drift increases from 0.03 to 0.21, and escalation curvature is positive (0.08), demonstrating super-linear authority growth as attacker capability increases. Defensive controls reduce the probability of compromise, boundary failures, and privilege drift, while introducing modest reductions in task utility and moderate latency overhead. By quantifying how design decisions impact amplification, instability, and escalation behavior, this study offers a structured evaluation of execution-coupled multi-agent AI systems and their security hardening. The main contributions of this work are:
\begin{itemize}
    \item \textbf{Amplification Analysis:} Quantifying how combining multiple agents under permissive execution rules amplifies the probability of system-level compromise and expands the impact surface.

    \item \textbf{Instability and Escalation Modeling:} Introducing quantitative metrics that link prompt injection to boundary instability, privilege drift, and super-linear privilege escalation dynamics.

    \item \textbf{Mitigation Evaluation Under Operational Constraints:} Evaluating defensive mechanisms using a counterfactual experimental methodology, measuring both security improvements and the resulting utility and latency trade-offs.
\end{itemize}

The remainder of the paper is organized as follows. Section~\ref{sec:related_work} reviews LLM-agent security and prompt injection. Section~\ref{Methodology} presents the methodology and formal system model. Section~\ref{Threat Model} defines the threat model and adversarial capabilities. Section~\ref{sec:results} reports empirical analysis and experimental results. Section~\ref{sec:discussion} discusses implications. Section~\ref{sec:threats} outlines threats to validity. Section~\ref{sec:limitations} describes limitations and future work. Section~\ref{sec:conclusion} concludes the paper.

\section{Related Work}
\label{sec:related_work}
This section reviews prior research on agent security and multi-agent LLM systems, focusing on prompt injection, aggregation risks, escalation dynamics, defense mechanisms, and supply-chain vulnerabilities.

\subsection{Prompt Injection and Multi-Agent Security}
Prompt injection attacks have gained increasing attention, particularly in tool-integrated and agentic systems. InjecAgent \cite{zhan2024injecagent} provides a systematic benchmark for assessing indirect prompt injection vulnerabilities in tool-augmented LLM agents. The study demonstrates that untrusted contextual inputs can manipulate tool invocation behavior, a risk amplified in environments with integrated tools and APIs. While InjecAgent focuses on attack success rates, it does not address how policies and boundary mechanisms impact overall system vulnerability. This limitation is particularly relevant for multi-agent systems such as OpenClaw, where agent aggregation can escalate risk, a phenomenon we quantify in this work. AgentDojo \cite{debenedetti2024agentdojo} operationalizes security evaluation in LLM agents by providing a dynamic testbed that measures attack success, defensive robustness, and benign utility across attack scenarios. While it demonstrates how adversarial content can compromise agent workflows, it lacks a formal examination of how inter-agent dynamics and policies contribute to systemic risk. Our work extends AgentDojo by incorporating multi-agent settings and quantifying associated risk amplification, which is critical for understanding the broader attack surface in multi-agent systems. CaMeL \cite{debenedetti2025camel} proposes a defense-by-design approach separating reasoning from execution to reduce prompt injection impact. While CaMeL offers architectural insights into mitigating injection, it primarily focuses on reasoning and tool layers rather than the boundary-defining components that we identify as critical sources of risk amplification. Our study extends this perspective by measuring the impact of policies on system-level risk, including parallel agent operation and permissive aggregation.

\subsection{Multi-Agent Aggregation Risks}
Coordination among diverse agents introduces unique challenges and potential security vulnerabilities \cite{wang2024survey}. Interactions among agents, especially when sharing execution substrates, can lead to coordination failures and resource contention. Prior studies do not explicitly examine how different aggregation policies directly amplify systemic security risks. In contrast, our work models these policies, particularly the permissive aggregation rule, which increases the likelihood of system compromise when any agent proposes a plausible action. This amplification is a key focus of our analysis. Further research includes \textit{Attacker Moves Second} \cite{nasr2025attackermovessecond}, which explores adaptive adversarial strategies bypassing static security measures. While it highlights exploitation of external attacks, it does not address internal vulnerabilities arising from agent interactions. Our study fills this gap by measuring how aggregation dynamics and privilege drift create new pathways for exploitation.

\subsection{Security Analysis and Escalation Dynamics}
Recent works examine escalation dynamics and privilege propagation in LLM-based agent systems. \textit{ToolHijacker} \cite{shi2025toolhijacker} introduces a vector by hijacking tool selection, highlighting the risk of injected prompts impacting system-level actions. However, it does not analyze the structural consequences of privilege escalation after a tool is invoked. In contrast, we quantify escalation using metrics such as privilege drift and attack surface entropy, providing a comprehensive view of escalation as adversarial capability increases. The study \textit{Privilege Escalation and Instability Analysis} \cite{gulyamov2025review} emphasizes privilege drift in multi-agent systems and proposes metrics for escalation under adversarial pressure. While insightful, it does not account for systemic impacts of boundary instability, a key driver of risk in self-hosted, execution-coupled systems such as OpenClaw. Our analysis introduces a more granular assessment of privilege instability in aggregation contexts, including positive curvature in escalation with increasing attacker capability.

\subsection{Defense Mechanisms and Systemic Risk Management}
\textit{Agent-Lock} \cite{hossain2025multiagentdefense} presents a multi-agent defense pipeline that detects and mitigates prompt injection using static analysis and dynamic monitoring. While informative, it does not focus on systemic vulnerabilities introduced by policies. Our work shows that mitigations that are impactful in isolation may still leave agents vulnerable when aggregated.
Design patterns for securing LLM agents \cite{search25designpatterns} provide guidance on reducing the likelihood of injection but do not specifically address interactions within shared execution environments. Our analysis fills this gap by assessing how aggregation policies mitigate or amplify systemic risks in the context of prompt injection and privilege escalation.

\subsection{Supply-Chain Risk in Multi-Agent Systems}
Supply-chain vulnerabilities are explored in \textit{Supply-Chain Risk Evaluation} \cite{tang2025massecurity}, which examines how malicious skills propagate in agent ecosystems. While providing a robust framework for skill-based vulnerabilities, it does not account for how injection-induced vulnerabilities interact with aggregation policies. Our work demonstrates that injection risks can propagate across the entire agent network when low-barrier execution is permitted.\\

The literature synthesis reveals that, while existing studies advance understanding of prompt injection and defensive mechanisms, there remains a gap in structural and systemic analysis of multi-agent execution coupling. Prior research often focuses on model-level vulnerabilities, static evaluation, and defense heuristics, but overlooks how policies and boundary instability shape overall system security. To address this, we employ formal metrics for amplification, CBI, and privilege escalation, along with a counterfactual evaluation for mitigation, to provide a detailed and actionable understanding of security risks in multi-agent LLM systems.

\section{Methodology}
\label{Methodology}
\begin{figure*}[t]
    \centering
    \includegraphics[width=0.80\linewidth]{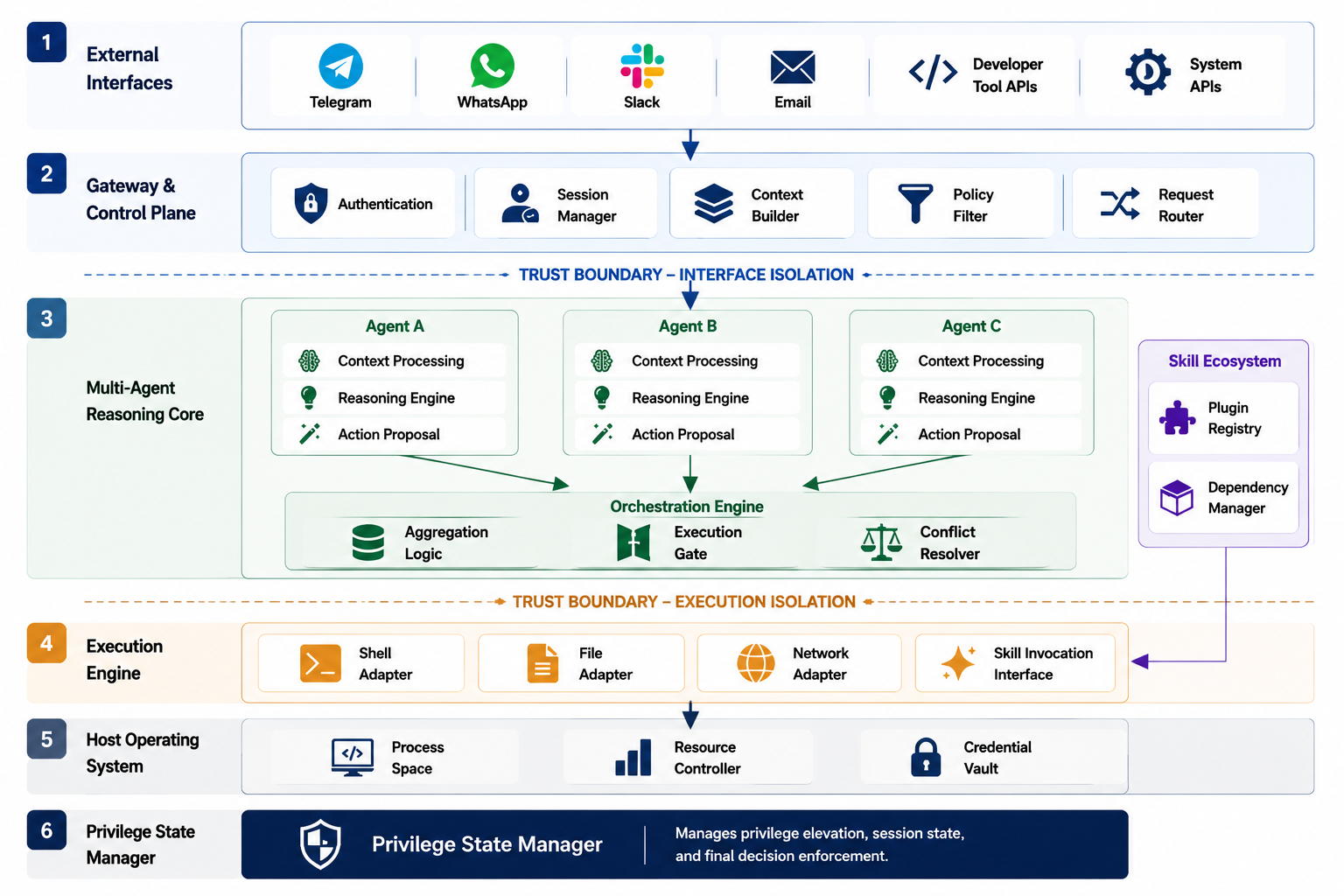}
    \caption{OpenClaw execution-coupled multi-agent system architecture.}
    \label{fig:openclaw_architecture}
\end{figure*}

We formalize the execution-coupled structure of OpenClaw, as its architectural composition directly determines the attack surface and privilege dynamics analyzed in this study. Figure~\ref{fig:openclaw_architecture} illustrates the layered system model serving as the structural foundation for our analysis. OpenClaw is organized as a pipeline in which contextual inputs enter through external interfaces, including messaging connectors, developer APIs, and system APIs. These inputs pass through a gateway and a control plane, which handle authentication, session management, context construction, and policy filtering, before crossing the first trust boundary separating untrusted external data from internal reasoning components.  Within this boundary, multiple LLM-based agents operate in parallel, each instantiated from the same LLM backend but with distinct configurations, depending on the experimental configuration. Each agent performs context processing, semantic reasoning, and action proposal generation. The resulting proposals are evaluated by a \textit{selection-and-execution controller}, which enforces aggregation logic, resolves conflicts, and applies execution policies. This controller is a structural amplification point: a single compromised agent may be sufficient to trigger system-level actions. Decision rules in the controller are task-specific and consistent across experiments, governing how individual agent outputs are translated into executable authority. A second trust boundary separates reasoning from execution. Moreover, the execution engine converts semantic outputs into concrete system operations using shell, file, network, and skill invocation adapters. These operations execute within the host OS, where process isolation, resource controls, and credential management define impactive privilege states. A privilege management component tracks authority transitions across execution stages. This decomposition clarifies how reasoning propagates into execution, establishing the foundation for formal risk modeling.    
We define an \textit{execution-coupled agentic system} as one in which model reasoning outputs are directly mapped to executable primitives, creating operational pathways through which adversarial inputs can propagate into privileged states. Unlike traditional LLM deployments, where outputs remain confined to textual responses, OpenClaw translates semantic outputs into actionable system operations. Three distinct LLM backends are evaluated: GPT-5.2, DeepSeek-R1, and Llama-4-Maverick. In each experimental configuration, agents instantiated from these models generate action proposals that are subsequently considered by the controller. We formalize the agent set as
\[
\mathbb{M} = \{m_1, m_2, m_3\} = \{\text{GPT-5.2}, \text{DeepSeek-R1}, \text{Llama-4-Maverick}\}.
\]
In single-agent configurations, OpenClaw executes proposals from one model per run. In multi-agent configurations, outputs from multiple models are aggregated through a \textit{selection policy}:
\[
\pi:\; \{O^{(m)}\}_{m\in\mathbb{M}} \rightarrow a^\star \in \mathcal{A}_{system},
\]
where $O^{(m)}$ denotes the output of model $m$, and $a^\star$ is the action selected for execution. If a proposal contains multiple executable actions, the controller applies the same policy to each action independently, thereby defining aggregation at the action level. This approach ensures clarity in risk attribution: compromising a single model and agent can impact the set of actions executed.    
The experimental methodology consists of four integrated layers: structural modeling, adversarial stimulation, runtime instrumentation, and statistical validation. These layers collectively feed into a unified risk evaluation methodology defined as
\[
\mathcal{F}_{risk} : (\mathcal{S}, \mathcal{A}_{attacker}, \mathcal{H}) \rightarrow \mathbb{R}^d,
\]
where $\mathcal{S}$ represents system configuration (agents, model instances, and execution substrate), $\mathcal{A}_{attacker}$ encodes attacker capabilities, and $\mathcal{H}$ denotes experimental assumptions, including task contexts and environmental conditions. The output vector comprises measurable security metrics, including attack surface magnitude, boundary failure probability, and privilege drift. Model-averaged and worst-case risk formulations account for multi-agent settings:
\[
\overline{\mathcal{F}}_{risk}(\mathcal{S}, \mathcal{A}_{attacker}, \mathcal{H}) 
= \frac{1}{|\mathbb{M}|} \sum_{m\in\mathbb{M}} 
\mathcal{F}_{risk}(\mathcal{S}, \mathcal{A}_{attacker}, \mathcal{H}; m),
\]
\[
\mathcal{F}^{\max}_{risk}(\mathcal{S}, \mathcal{A}_{attacker}, \mathcal{H}) 
= \max_{m\in\mathbb{M}} 
\mathcal{F}_{risk}(\mathcal{S}, \mathcal{A}_{attacker}, \mathcal{H}; m).
\]
The model-averaged formulation reflects expected risk across heterogeneous backends, while the worst-case formulation captures conservative exposure under the most vulnerable agent. This dual perspective ensures that risk estimates reflect structural and aggregation impacts rather than incidental variations in model behavior.    

\subsection{Explicit Attacker Model}
We represent attacker capability as a four-dimensional vector:
\[
\mathcal{A}_{attacker} = (N, P, S, L),
\]
where each component reflects a distinct domain of potential impact and is normalized to the interval $[0,1]$ to maintain comparability across dimensions. Normalization allows consistent scaling when combining heterogeneous factors such as network reach, message manipulation, skill exploitation, and local privilege.  
\begin{itemize}
    \item $N$ captures network reachability, from restricted local access to public Internet exposure.
    \item $P$ represents the ability to inject and shape messages consumed by agents, e.g., crafted inputs via Slack and Telegram that the system ingests.
    \item $S$ denotes impact over skills and extensions, such as publishing a malicious skill and persuading a user to activate it.
    \item $L$ reflects local system access, ranging from standard user privileges to elevated host-level authority.
\end{itemize}
Furthermore, these dimensions define a bounded hypercube in $\mathbb{R}^4$, ensuring that attacker strength remains finite and realistic. A partial order is defined over capability vectors:
\[
(N_0, P_0, S_0, L_0) \preceq (N_1, P_1, S_1, L_1)
\]
if each component of the first vector represents less capability than the corresponding component of the second. For example, lower $L$ indicates fewer privileges, and lower $P$ reflects weaker control over message injection. Under this ordering, attacker $\mathcal{A}_n \succeq \mathcal{A}_0$ implies at least as much potential impact across all dimensions, supporting monotonic reasoning: if a vulnerability exists under $\mathcal{A}_0$, it persists under any attacker $\mathcal{A}_n$. This methodology enables controlled escalation analysis in experiments by incrementally increasing each dimension and observing which transitions violate trust boundaries and execution constraints.
In multi-agent settings, attackers may exploit policy asymmetries across models. Let $Succ(m)$ indicate whether agent $m$ produces an unauthorized action under a given injection scenario. Under a permissive aggregation policy, where execution proceeds if any plausible action is proposed:
\[
Succ_{\pi} = \mathbb{1}\left(\exists m \in \mathbb{M} : Succ(m) = 1 \right),
\]
A single compromised agent is sufficient to affect system behavior. Under a conservative policy, such as a consensus requiring a majority of agents to agree:
\[
Succ_{\pi} = \mathbb{1}\left(\sum_{m \in \mathbb{M}} Succ(m) \geq \left\lceil \frac{|\mathbb{M}|}{2} \right\rceil \right),
\]
Coordinated compromise across multiple agents is required. In our experiments with $|\mathbb{M}| = 3$, this threshold corresponds to either 2 or 3 agents generating unauthorized actions.
For comparative analysis, we define a scalar norm over the attacker space:
\[
\|\mathcal{A}_{attacker}\| = \omega_N N + \omega_P P + \omega_S S + \omega_L L,
\]
where weights $\omega_N, \omega_P, \omega_S, \omega_L$ are fixed prior to experimentation to reflect realistic enterprise risk priorities. This scalar enables one-dimensional sensitivity curves that characterize how the attack surface magnitude and privilege dynamics evolve as adversarial capability increases.  Attacker injection scenarios are pre-defined and fixed within the experimental setup, ensuring repeatable conditions across different models and policy configurations. The attacker model itself is not an LLM; it represents controlled adversarial inputs systematically applied to evaluate system robustness.

\subsection{Formal System Model}
\label{Formal System Model}
We represent OpenClaw as the tuple
\[
\mathcal{S} = (I, G, M, E, K, O, \Pi),
\]
where each component corresponds to a security-relevant module in the execution pipeline:
\begin{itemize}
    \item $I$: input interface layer, including messaging connectors and ingestion channels.
    \item $G$: gateway and controller, responsible for authentication, routing, and session management.
    \item $M$: reasoning module instantiated from LLM backends (GPT-5.2, DeepSeek-R1, Llama-4-Maverick). $M$ is probabilistic, while all other modules are deterministic.
    \item $E$: execution layer, comprising system-facing tools such as shell, filesystem, and network adapters.
    \item $K$: skill ecosystem, representing installable plugins and their dependencies.
    \item $O$: host operating system layer, capturing processes, files, and stored secrets (e.g., API keys, session tokens, SSH keys).
    \item $\Pi$: privilege state space, formalizing how authority is represented and evolves across transitions.
\end{itemize}
Decoding parameters for $M$ (e.g., temperature, top-$p$, seed) are fixed during experiments to control stochasticity and ensure reproducibility. In multi-agent configurations, the system extends to
\[
\mathcal{S}_{MA} = (\mathcal{S}, \mathbb{M}, \pi),
\]
where $\mathbb{M}$ is the set of agent instances and $\pi$ is the action aggregation policy. This extension does not alter the underlying modules but makes explicit that $M$ may be instantiated multiple times and that $\pi$ governs how proposed actions are combined. We define system state privilege at $S_j$ as
\[
Priv(S_j) = \sigma(C_j, A_j),
\]
where $C_j$ is the set of accessible credentials (e.g., tokens, API keys, session cookies, SSH keys), and $A_j$ is the set of executable action primitives (e.g., file operations, shell commands, network calls, skill installation). Here, $E$ and $O$ define the set of available actions and credential sources, while $\Pi$ captures the overall authority state. The mapping $\sigma(\cdot)$ converts these sets into a scalar value for quantitative comparison:
\[
Priv(S_j) = \alpha |C_j| + \beta \log(|A_j| + 1),
\]
where $\alpha$ and $\beta$ are weights reflecting the relative impact of credential access and action breadth. The logarithmic term models diminishing returns: the first high-impact action substantially increases privilege, whereas additional low-impact actions contribute marginally. The concavity condition
\[
\frac{d^2 Priv}{d|A|^2} < 0
\]
ensures sublinear growth, preventing artificial inflation of authority in environments with large tool registries. Constants $\alpha$ and $\beta$ are selected based on the relative risk importance of credentials versus action primitives in the system. Per-transition privilege gain is defined as
\[
\Delta Priv_j = Priv(S_{j+1}) - Priv(S_j),
\]
where each transition $S_j \rightarrow S_{j+1}$ corresponds to a stage in the OpenClaw pipeline (e.g., message ingestion, LLM reasoning, tool invocation, skill loading). Cumulative Privilege Drift (CPD) over a trace of length $n$ is
\[
CPD = \sum_{j=1}^{n} \Delta Priv_j.
\]
A positive CPD without explicit policy authorization constitutes a boundary violation, thereby enabling unintended gains in authority. To enable comparison across environments, we define normalized privilege drift:
\[
NPD = \frac{CPD}{Priv(S_0)},
\]
scaling cumulative drift relative to the initial privilege. This ensures that identical absolute privilege gains are interpreted in proportion to the baseline authority, providing a consistent metric for evaluating the impact of escalation and injection across different system configurations.

\subsection{Attack Surface Quantification}
We quantify the OpenClaw attack surface as
\[
\mathcal{A} = \sum_{i=1}^{n} P_i \cdot I_i \cdot \gamma_i \cdot \lambda_i,
\]
where the system is decomposed into $n$ exposure vectors $e_i$, representing security-relevant interfaces such as gateway ports, skill-installation endpoints, and tool-execution interfaces. Each component of the summation is defined as follows:
\begin{itemize}
    \item $P_i$ is the empirically measured probability of successful exploitation of $e_i$ under a specified attacker capability configuration. It is estimated through repeated adversarial trials in which controlled injections and manipulations are applied, with success observed.
    \item $I_i$ denotes the impact of a successful exploit, quantified as the fraction of system resources, credentials, and operational objects reachable through the exposure.
    \item $\gamma_i$ captures privilege amplification, measured as the ratio of system privilege immediately after exploitation to the privilege prior to exploitation. This is computed using the per-transition privilege function $Priv(S_j)$, as defined in the formal system model, with experimental traces that record pre- and post-exploit states.
    \item $\lambda_i$ reflects persistence, representing the expected duration an exposure remains exploitable before remediation and automatic updates neutralize the threat.
\end{itemize}
All components are normalized to $[0,1]$ to ensure boundedness and comparability across different exposure vectors:
\[
0 \leq P_i, I_i, \gamma_i, \lambda_i \leq 1.
\]
This guarantees that $\mathcal{A} \le n$, providing a stable metric across system versions. Increasing $n$ introduces additional exposure vectors while maintaining comparability through normalized sub-scores.
We define system-level risk as the combination of probability, impact, privilege amplification, and persistence across all exposure vectors. Risk severity is impacted by the magnitude of $P_i$, the criticality of impacted resources ($I_i$), the potential for privilege escalation ($\gamma_i$), and the duration of exploitability ($\lambda_i$). To characterize the distribution of risk across vectors, we define structural entropy:
\[
H_A = - \sum_{i=1}^{n} p_i \log p_i, \quad
p_i = \frac{P_i I_i \gamma_i \lambda_i}{\sum_{j=1}^{n} P_j I_j \gamma_j \lambda_j}.
\]
Low entropy indicates concentrated risk in a few dominant exposures, suggesting targeted hardening is impactive. High entropy indicates distributed risk across multiple moderate exposures, motivating layered defensive strategies. We further define attack surface curvature:
\[
\kappa_A = \frac{d^2 \mathcal{A}}{d\|\mathcal{A}_{attacker}\|^2},
\]
which measures the acceleration of risk growth as attacker capability increases. Positive curvature ($\kappa_A > 0$) indicates super-linear escalation, where incremental increases in capability lead to disproportionately larger exposure, e.g., when expanded network reach and supply-chain impact unlock multiple dependent attack paths. In multi-agent configurations, the impactive attack surface is policy-dependent. Let $\mathcal{A}^{(m)}$ denote the attack surface when agent $m$ is active. The policy-conditioned attack surface is
\[
\mathcal{A}_{\pi} =
\begin{cases}
1 - \prod_{m\in\mathbb{M}} (1-\mathcal{A}^{(m)}) & \text{(permissive aggregation)},\\[2mm]
\sum_{m\in\mathbb{M}} w_m\,\mathcal{A}^{(m)} & \text{(weighted routing)}.
\end{cases}
\]
The first case reflects permissive aggregation, in which executing any single agent's proposal increases system-level exposure. The second case models routing-based configurations in which each agent is invoked according to the weight $w_m$, reflecting deployment preferences such as the frequency of agent invocation and the criticality of assigned roles. These weights allow risk exposure to be modulated in accordance with system design priorities.

\subsection{Attack Surface and Risk Distribution}
The OpenClaw attack surface is modeled as the sum of multiple exposure vectors, each representing a potential entry point for adversarial actions:
\[
\mathcal{A} = \sum_{i=1}^{n} P_i \cdot I_i \cdot \gamma_i \cdot \lambda_i,
\]
where \(P_i\) denotes the empirically measured likelihood of exploitation for the $i$-th exposure vector, \(I_i\) measures the impact of exploitation, \(\gamma_i\) quantifies privilege amplification, and \(\lambda_i\) represents the persistence of the exposure. All factors are normalized to the interval [0,1] to ensure comparability across different vectors. To characterize how risk is distributed across these vectors, we define the attack surface entropy:
\[
H_A = - \sum_{i=1}^{n} p_i \log p_i, \quad
p_i = \frac{P_i I_i \gamma_i \lambda_i}{\sum_{j=1}^{n} P_j I_j \gamma_j \lambda_j}.
\]
Low entropy indicates that a small number of vulnerabilities disproportionately contribute to the overall risk, making targeted mitigation impactive. High entropy implies that risk is distributed across many vectors, necessitating layered defensive measures. Measuring entropy provides a deeper understanding of the system's risk profile, helping prioritize mitigation strategies and allocate defensive resources impactively.  This formulation complements the previously defined metrics by linking exposure probability, impact, privilege amplification, and persistence to a quantitative assessment of risk magnitude and distribution across OpenClaw's execution-coupled architecture.

\subsection{Prompt Injection Evaluation}
We model prompt injection as the transformation
\[
\tilde{P} = P \oplus \delta,
\]
where \(P\) denotes a benign task prompt and associated user message thread, and \(\delta\) represents an adversarial payload generated by the attacker. In our experimental setup, \(\delta\) is derived from predefined attack scenarios that simulate malicious users. The operator \(\oplus\) denotes concatenation using the same formatting and message serialization procedure that OpenClaw employs when constructing LLM inputs, ensuring that injection evaluation reflects realistic system behavior rather than abstract text concatenation. To quantify the impact of prompt injection on privilege escalation, we define the \textit{Injection Amplification Factor} (IAF):
\[
IAF = \frac{\Delta Priv_{injected}}{\Delta Priv_{benign}},
\]
where \(\Delta Priv_{benign}\) is the privilege change observed under the benign prompt \(P\), and \(\Delta Priv_{injected}\) is the privilege change under the adversarial prompt \(\tilde{P}\). A value of \(IAF > 1\) indicates that the injected prompt caused a greater privilege gain than the benign task alone. This interpretation relies on the scalar privilege function \(Priv(S_j)\) defined in the Formal System Model \ref{Formal System Model}, which captures both credential access and executable action breadth. Comparing \(\Delta Priv_{injected}\) to \(\Delta Priv_{benign}\) ensures that the measurement reflects only the additional authority gained due to the injection, independent of task-specific baseline privilege requirements.  
This metric enables systematic evaluation of how prompt injections propagate through reasoning-to-execution pathways and quantifies their impact on exploitability under realistic multi-agent deployment conditions.

\paragraph{Semantic Sensitivity}
We define semantic sensitivity as
\[
SS = \frac{\partial Priv}{\partial \|\delta\|},
\]
measuring the rate at which system privilege changes as the magnitude of an adversarial payload increases. In practice, \(\|\delta\|\) is parameterized by the payload size, instruction density, and the intensity of structured perturbation. High semantic sensitivity indicates that small textual and structured modifications in the input can induce large shifts in privilege, reflecting a fragile coupling between semantic reasoning and execution authority. In the multi-agent configuration, injection impacts are evaluated at both the individual-agent and policy-aggregation levels. Let \(IAF^{(m)}\) denote the IAF observed for agent \(m\), instantiated from a specific LLM backend. Let $IAF^{(m)}$ denote the Injection Amplification Factor observed for agent $m$, where each agent is a distinct instance in the system. In this context, $m$ always refers to a single agent, independent of which LLM model it uses. Under permissive selection policies, where execution proceeds if any agent proposes an action, the impactive amplification is determined by the most vulnerable agent:
\[
IAF_{\pi}^{\max} = \max_{m \in \mathbb{M}} IAF^{(m)}.
\]
Under consensus-based policies, such as majority agreement for tool invocation, amplification depends on the number of agents simultaneously induced to generate unauthorized actions. In this scenario, we empirically estimate the policy-level injection rate \(PI_{rate,\pi}\) and compute:
\[
IAF_{\pi} = \frac{\Delta Priv_{\text{injected},\pi}}{\Delta Priv_{\text{benign},\pi}},
\]
where privilege changes are measured after aggregation according to policy \(\pi\). This formulation directly links injection robustness to aggregation rules and supports systematic evaluation of mitigation strategies, including consensus gating and execution veto mechanisms, within a unified quantitative framework.

\subsection{Extended Security Evaluation Algorithm}
To operationalize the structural and quantitative metrics defined above, we implement a deterministic security evaluation procedure, summarized in Algorithm~\ref{alg:openclaw_eval}. This procedure systematically evaluates exposure vectors, privilege escalation, trust boundary violations, and skill-level risks under a fixed attacker capability configuration.
\begin{algorithm}[t]
\caption{Deterministic OpenClaw Security Evaluation}
\label{alg:openclaw_eval}
\footnotesize
\textbf{Input:} System state $\mathcal{S}$, agent $m$, attacker $\mathcal{A}_{attacker}$ \\
\textbf{Output:} $(\mathcal{R}, \mathcal{A}, BR)$
\begin{algorithmic}[1]
\State SeedRandom(42)
\State $Priv_0 \leftarrow$ MeasurePrivilege($\mathcal{S}$)
\State Assert(Consistency($\mathcal{H}$) == True)
\For{each exposure vector $e_i$ defined in Section~\ref{methodology_attack_surface_revised}}
    \State SimulateAttack($\mathcal{S}, e_i, \mathcal{A}_{attacker}$)
    \State $\Delta Priv \leftarrow$ MeasurePrivilege($\mathcal{S}$) - $Priv_0$
    \State UpdateMetrics($e_i$, $\Delta Priv$)
\EndFor
\For{each trust boundary $B_j$ introduced in Section~\ref{Methodology}}
    \State StressTestBoundary($\mathcal{S}, B_j$)
    \State ComputeBoundaryMetrics($B_j$)
\EndFor
\For{each skill $k$ in the skill ecosystem $K$}
    \State StaticAnalyze($k$)
    \State DynamicExecuteSandbox($k$)
    \State ComputeRisk($k$)
\EndFor
\State SimulateGatewayCompromise($\mathcal{S}$)
\State AggregateMetrics()
\State \Return $(\mathcal{R}, \mathcal{A}, BR)$
\end{algorithmic}
\end{algorithm}
The function `SimulateAttack` applies adversarial payloads from the Attacker Model to each exposure vector and measures the induced changes in system privilege. `StressTestBoundary` evaluates whether privilege escalation and injection propagation cross trust boundaries between untrusted input, reasoning, and execution layers. The combination of static analysis and sandboxed dynamic execution assesses skill-level risks, ensuring that installed plugins do not introduce unauthorized authority. Although LLMs exhibit stochastic behavior, setting a fixed random seed and standardizing decoding parameters ensures consistent experimental conditions. Under identical system configurations, attacker capabilities, and seed values, the sequence of simulation steps and resulting metrics remains consistent, with only minor measurement noise introduced by runtime instrumentation. All data, testing results, and scripts are provided in a replication package to support reproducibility, in line with standard practices in software engineering research. The evaluation iterates over finite sets of exposure vectors, trust boundaries, and skills, making the procedure computationally tractable even as the skill ecosystem grows. In multi-agent configurations, the algorithm is executed independently for each agent to obtain per-agent risk vectors, and then aggregated to compute policy-level metrics. This separation enables the identification of vulnerabilities specific to individual models, as well as those arising from aggregation, thereby aligning the measurement process with the structural abstraction of execution-coupled multi-agent architectures.

\subsection{Mitigation Design Space and Hardening Operators}
The previous subsections characterize architectural risk in OpenClaw. The second objective of this study is to introduce mitigations and evaluate their impactiveness under identical adversarial conditions. We formalize each mitigation as a hardening operator that transforms a baseline system into a defended instance:
\[
\mathcal{S}^{(\mathcal{H})} = \mathcal{H}(\mathcal{S}),
\]
where \(\mathcal{H}\) denotes a specific mitigation, and \(\mathcal{S}^{(\mathcal{H})}\) represents the resulting hardened deployment. This operator-based formulation embeds defensive mechanisms directly within the same quantitative risk framework, enabling measurable comparison rather than purely qualitative assessment. We define the mitigation space as
\[
\mathbb{H} = \{\mathcal{H}_1, \mathcal{H}_2, \dots, \mathcal{H}_q\},
\]
By assigning each element to a reproducible control applicable to OpenClaw, the mitigation space $\mathbb{H}$ includes configuration-level patches such as policy definitions, container profiles, gateway flags, and skill registry constraints, thereby ensuring systematic deployability.  
To align defensive mechanisms with OpenClaw's trust boundaries, mitigations are organized into five families: 1) Execution Policy Hardening enforces gating and step-up confirmation for high-impact tools. High-impact tools are those that can directly modify system state or access sensitive credentials, such as executing shell commands or installing software. Step-up confirmation may require explicit approval from multiple agents and user acknowledgment before execution. 2) Prompt-Channel Isolation separates untrusted message contexts from critical reasoning pipelines and applies structured injection filtering, including pattern-based sanitization, context-length limits, and controlled parsing across different input channels (e.g., Slack, Telegram) to prevent propagation of malicious payloads. 3) Skill Ecosystem Governance applies allowlists, signature verification, and dependency constraints to prevent unverified or malicious plugins from being installed and executed. 4) Runtime Containment uses confinement mechanisms, including seccomp profiles, AppArmor policies, least-privilege containerization, and filesystem isolation, to limit the potential impact of an exploited process and restrict process capabilities. 5) Network Exposure Minimization enforces secure-by-default service bindings, authenticated gateways, and network segmentation to reduce the attack surface exposed to external and lateral movement.
In multi-agent deployments, the aggregation policy $\pi$ serves as a hardening mechanism. A conservative, consensus-gated execution rule requires agreement across multiple agents before high-impact actions are permitted. High-impact actions can change system state or elevate privileges (e.g., executing a shell command or installing a skill), whereas low-impact actions are read-only. Formally:
\[
Execute(a) = \mathbb{1}\left(\sum_{m\in\mathbb{M}} \mathbb{1}[a \in Proposal(m)] \ge \tau_\pi \right),
\]
where $\tau_\pi$ is a predefined voting threshold. This mechanism transforms agent diversity into a security control, requiring agreement on proposals before execution. impactiveness is evaluated using the same counterfactual protocol as for other mitigations, to ensure consistent comparisons across defense families.

\subsection{Mitigation Attribution, Counterfactual Evaluation, and Acceptance Criteria}
Mitigation design in OpenClaw proceeds through three interconnected stages: (1) attribution of individual controls, (2) counterfactual measurement of their causal impact, and (3) enforcement of multidimensional acceptance criteria. Moreover, these stages establish a rigorous, quantitative methodology for evaluating defensive mechanisms rather than relying on qualitative assessment. When multiple mitigations are combined into a composite defense profile, it is necessary to determine each component's contribution. Let \(\mathcal{H}_{a+b}\) denote a composite mitigation composed of components \(\mathcal{H}_a\) and \(\mathcal{H}_b\). We quantify interaction impacts using a synergy measure:
\[
Synergy(a,b) = \Delta \mathcal{A}_{a+b} - \big( \Delta \mathcal{A}_a + \Delta \mathcal{A}_b \big),
\]
where \(\Delta \mathcal{A}_*\) indicates the reduction in attack surface due to the specified mitigation. A positive synergy value indicates complementary protection, meaning the combined mitigation reduces exposure beyond the sum of individual impacts. A negative value indicates overlap, reflecting diminishing returns when multiple controls address the same vulnerabilities. This attribution methodology supports principled construction of hardened deployment profiles. To isolate the causal impact of each mitigation, we adopt a counterfactual experimental design. For every attacker profile \(\mathcal{A}_{attacker}\) and fixed execution seed, we execute the evaluation pipeline on both the baseline system and its hardened counterpart:
\begin{align}
\mathcal{R}_0 &= \mathcal{F}_{risk}(\mathcal{S}, \mathcal{A}_{attacker}, \mathcal{H}), \\
\mathcal{R}_m &= \mathcal{F}_{risk}(\mathcal{S}^{(m)}, \mathcal{A}_{attacker}, \mathcal{H}),
\end{align}
where \(\mathcal{R}_0\) and \(\mathcal{R}_m\) are \(d\)-dimensional vectors of measured security metrics. The mitigation impact is computed as
\[
\Delta \mathcal{R}_m = \mathcal{R}_m - \mathcal{R}_0.
\]
For risk-related components, including attack surface magnitude \(\mathcal{A}\), CPD, boundary failure probability \(P(B_j^{fail})\), and IAF, negative values in \(\Delta \mathcal{R}_m\) indicate reduced exposure. Because the attacker's capabilities, environmental configuration, and random seeds are held constant across paired evaluations, observed differences can be directly attributed to the applied hardening operator. This counterfactual approach eliminates confounding factors and ensures consistent comparability across mitigation strategies.
Mitigation acceptance is defined by multidimensional improvement rather than optimization of a single metric. Primary acceptance conditions are:
\[
\Delta \mathcal{A}_m < 0, \quad
\Delta CPD_m < 0, \quad
\Delta IAF_m \le 0,
\]
ensuring reductions in attack surface magnitude, CPD, and injection amplification. Additionally, boundary-level strengthening is required:
\[
\Delta P(B_j^{fail})_m < 0 \quad \forall j \in \{1, \dots, J\}.
\]
Every modeled trust boundary must show a decrease in failure probability after mitigation. These constraints prevent displacement impacts, where improvement along one dimension is offset by degradation along another. A mitigation is therefore accepted only if it produces coherent, systemic hardening across architectural layers.

\subsection{Security-Utility Trade-off and Operational Constraints}
Because OpenClaw is designed for productive automation, mitigations must preserve operational viability. Hardening mechanisms are therefore evaluated under explicit overhead constraints:
\[
\Omega = \{ L \le L_{\max},\; E_{exec} \le E_{\max},\; U_{task} \ge U_{\min} \},
\]
where:  
\begin{itemize}
    \item $L$ denotes end-to-end tool invocation latency.  
    \item $E_{exec}$ represents per-execution energy consumption, captured by the runtime monitoring layer.  
    \item $U_{task}$ measures benign-task utility, defined as the fraction of non-adversarial tasks completed successfully without unjustified blocking.
\end{itemize}
Mitigation selection is formalized as the constrained optimization problem:
\[
\min_{\mathcal{H} \in \mathbb{H}} \; \mathcal{A}(\mathcal{S}^{(\mathcal{H})})
\quad \text{s.t.}\quad \mathcal{S}^{(\mathcal{H})} \in \Omega,
\]
capturing the objective of minimizing measurable attack surface while satisfying operational constraints. This ensures that security improvements remain compatible with deployment requirements, preventing degradation of functional performance.
\begin{figure*}[t]
    \centering
    \includegraphics[width=0.75\linewidth]{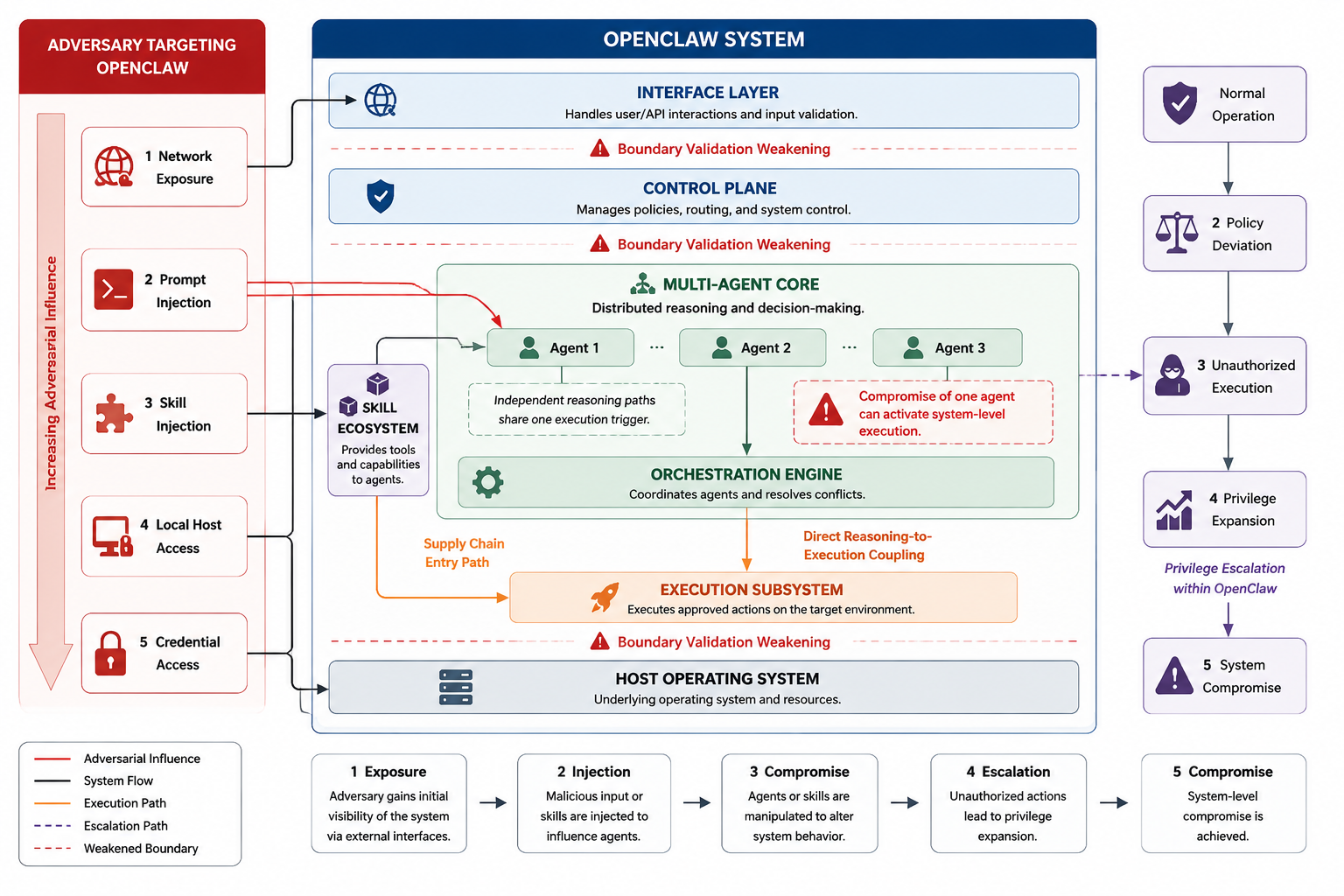}
    \caption{Adversarial escalation paths in OpenClaw, illustrating how network exposure, prompt manipulation, skill injection, local access, and credential compromise propagate through the multi-agent core and execution layers, enabling privilege amplification and potential system compromise.}
    \label{fig:threat_model}
\end{figure*}

\subsection{Mitigation Verification Algorithm}
To evaluate mitigations under the constrained security–utility objective, we implement a counterfactual verification procedure summarized in Algorithm~\ref{alg:mitigation_verification}. The algorithm performs paired baseline and defended evaluations for each attacker profile, computes risk deltas, and enforces operational feasibility constraints \(\Omega\) before ranking mitigations.
\begin{algorithm}[t]
\caption{Counterfactual Verification of OpenClaw Mitigations}
\label{alg:mitigation_verification}
\footnotesize
\textbf{Input:} Baseline system \(\mathcal{S}\), mitigation set \(\mathbb{H}\), attacker profiles \(\mathbb{A}\), experimental assumptions \(\mathcal{H}\) \\
\textbf{Output:} Mitigation ranking by risk reduction under operational constraints \(\Omega\)
\begin{algorithmic}[1]
\State SeedRandom(42)
\For{each attacker profile \(\mathcal{A}_{attacker} \in \mathbb{A}\)}
    \State \(\mathcal{R}_0 \leftarrow \mathcal{F}_{risk}(\mathcal{S}, \mathcal{A}_{attacker}, \mathcal{H})\)
    \For{each mitigation \(\mathcal{H}_i \in \mathbb{H}\)}
        \State \(\mathcal{S}^{(\mathcal{H}_i)} \leftarrow \mathcal{H}_i(\mathcal{S})\)
        \State \(\mathcal{R}_i \leftarrow \mathcal{F}_{risk}(\mathcal{S}^{(\mathcal{H}_i)}, \mathcal{A}_{attacker}, \mathcal{H})\)
        \State \(\Delta \mathcal{R}_i \leftarrow \mathcal{R}_i - \mathcal{R}_0\)
        \State \((L, E_{exec}, U_{task}) \leftarrow \text{MeasureOverhead}(\mathcal{S}^{(\mathcal{H}_i)})\)
        \If{\((L, E_{exec}, U_{task}) \in \Omega\)}
            \State StoreResult(\(\mathcal{A}_{attacker}, \mathcal{H}_i, \Delta \mathcal{R}_i\))
        \EndIf
    \EndFor
\EndFor
\State RankMitigationsBy(\(\Delta \mathcal{A}\), \(\Delta CPD\), \(\Delta P(B^{fail})\), \(\Delta IAF\))
\State \Return ranked mitigations and per-attacker deltas
\end{algorithmic}
\end{algorithm}
The procedure enforces a controlled before-and-after comparison under fixed attacker capability, deployment configuration, and random seed. Because environmental conditions remain invariant across paired evaluations, the resulting \(\Delta \mathcal{R}_i\) values represent the causal impact of each hardening operator rather than incidental measurement variation. This methodology ensures a consistent, quantitative assessment of mitigation effectiveness while accounting for operational constraints and the impacts of multi-agent aggregation.

\section{Threat Model}
\label{Threat Model}

Figure~\ref{fig:threat_model} illustrates the adversarial landscape targeting OpenClaw. The attack surface expands as adversarial capability increases, beginning with network reachability and prompt manipulation, extending to skill injection and local host access, and potentially culminating in credential compromise. Because OpenClaw directly connects semantic reasoning to privileged execution, compromise of a single agent can propagate through aggregation and execution subsystems, producing system-level impacts.  
The purpose of this threat model is to formally describe how adversarial control translates into exposure, privilege amplification, and eventual system compromise. We define a threat-mapping function:
\[
\mathcal{T} : (\mathcal{S}, \mathcal{A}_{attacker}) \rightarrow \mathcal{R}_{threat},
\]
where \(\mathcal{S}\) denotes the system configuration, \(\mathcal{A}_{attacker}\) denotes adversarial capability, and \(\mathcal{R}_{threat}\) is a structured risk state. Rather than a single scalar, \(\mathcal{R}_{threat}\) comprises measurable quantities aligned with the evaluation, including expected privilege drift, boundary-failure probabilities, supply-chain risk, compromise probability, and escalation likelihood.

\subsection{Adversary Capability Model}
The adversary is modeled as a five-dimensional capability vector:
\[
\mathcal{A}_{attacker} = (N, P, S, L, C),
\]
representing network access (\(N\)), prompt manipulation capability (\(P\)), skill injection ability (\(S\)), local privilege level (\(L\)), and credential acquisition capability (\(C\)). The additional dimension \(C\) explicitly captures the potential for credential compromise, complementing the four dimensions of the previously introduced attacker model and enabling more precise evaluation of privilege escalation pathways.  Each dimension is normalized to \([0,1]\) to ensure bounded adversarial strength. To support sensitivity and escalation analysis, we define a scalar adversary norm:
\[
\|\mathcal{A}_{attacker}\| = \sum_{i=1}^{5} \omega_i a_i,
\]
where weights \(\omega_i\) reflect deployment-specific criticality and are fixed prior to experimentation. This scalarization enables a systematic study of how risk evolves as adversarial capability increases. We further define a dominance relation:
\[
\mathcal{A}_1 \preceq \mathcal{A}_2 \iff a_{1i} \le a_{2i} \quad \forall i,
\]
which induces a capability lattice. This structure enables controlled escalation experiments in which one capability dimension is increased while others remain fixed, supporting monotonic reasoning: if a vulnerability exists under \(\mathcal{A}_1\), it persists under any stronger adversary \(\mathcal{A}_2 \succeq \mathcal{A}_1\).

\subsection{Capability-Conditioned Attack Progression}
We model an attack as a sequence of state transitions:
\[
S_0 \xrightarrow{\alpha_1} S_1 \xrightarrow{\alpha_2} \dots \xrightarrow{\alpha_k} S_k,
\]
where each adversarial action \(\alpha_i\) is constrained by the adversary capability vector \(\mathcal{A}_{attacker}\).  
An attack is considered successful if the cumulative privilege gain exceeds a meaningful operational threshold \(\theta\):
\[
Priv(S_k) - Priv(S_0) > \theta.
\]
The threshold \(\theta\) is empirically determined from observed privilege dynamics during benign task execution and the system’s baseline operational risk. Specifically, \(\theta\) is set to exceed typical stochastic fluctuations in \(\Delta Priv\) under benign conditions, ensuring that only escalations with operational significance, such as shell access, skill installation, and credential compromise, are counted as successful attacks. The empirical probability of escalation under a given attacker capability configuration is therefore defined as:
\[
P_{esc} = P\big(Priv(S_k) > Priv(S_0) + \theta \mid \mathcal{A}_{attacker}\big).
\]
This formulation ties attack success to operationally meaningful privilege amplification rather than arbitrary metric variations, ensuring that measured escalation reflects actionable impact within OpenClaw’s multi-agent, execution-coupled environment.

\subsection{Semantic-to-Execution Exploitation}
The primary security risk in OpenClaw arises from the mapping between LLM outputs and system actions:
\[
\Phi : \mathcal{O}_{semantic} \rightarrow \mathcal{A}_{system},
\]
where semantic artifacts( e.g., plans, prompts, and tool call proposals) are transformed into executable system operations. An adversarial perturbation \(\delta\), embedded within a benign prompt \(P\), aims to maximize the resulting privilege state:
\[
\max_{\delta} \; Priv(\Phi(M(P+\delta))),
\]
subject to the enforcement of execution policies and trust boundaries.
We quantify the sensitivity of this semantic-to-execution mapping using the \textit{Semantic Exploitability Coefficient} (SEC):
\[
SEC = \frac{\Delta Priv_{semantic}}{\|\delta\|},
\]
where \(\|\delta\|\) is the magnitude of the adversarial perturbation, normalized by token count and instruction density. A high SEC indicates that small, carefully crafted textual manipulations produce disproportionately large privilege gains, highlighting fragile coupling between semantic reasoning and execution authority. This metric complements the IAF and policy-conditioned evaluation, providing a fine-grained view of how semantic perturbations propagate into system-level exploitation in a multi-agent, execution-coupled environment.

\subsection{Trust Boundary Violations}
Each architectural boundary \(B_j\) in OpenClaw is associated with a validation predicate \(\Theta(\cdot)\), which enforces policy and trust constraints. A boundary violation occurs when an input crossing the boundary is accepted (\(\Theta(x)=1\)) yet results in an unauthorized privilege increase:
\[
Priv(T(x)) > Priv(x),
\]
where \(T(x)\) denotes the system state after processing the input \(x\).
Boundary failure probability is estimated empirically:
\[
P(B_j^{fail}) = \frac{\text{number of violating crossings}}{\text{total tested crossings}}.
\]
To account for varying boundary criticality, we define the CBI metric:
\[
CBI = \sum_{j} P(B_j^{fail}) \cdot Impact(B_j),
\]
which ensures that rare, high-impact boundary failures are properly weighted relative to frequent, low-impact failures. Here, \(Impact(B_j)\) can be quantified using the potential privilege amplification, access to high-value credentials, and criticality of actions permitted across the boundary.
In multi-agent and execution-coupled contexts, CBI captures how violations at any single boundary can propagate through aggregation and execution pipelines, enabling system-level compromise even if most other boundaries remain intact. This metric integrates directly with attack surface, semantic sensitivity, and CPD to provide a holistic assessment of architectural resilience.

\subsection{Supply-Chain and Network Threats}

Within the skill ecosystem, a malicious skill \(k^*\) produces a payoff defined as:
\[
U(k^*) = Priv(S_{post}) - Priv(S_{pre}) - DetectionCost(k^*),
\]
where \(Priv(S_{post})\) and \(Priv(S_{pre})\) denote the privilege states after and before installation, and \(DetectionCost(k^*)\) captures the defensive detection and containment overhead.

Systemic supply-chain risk is then quantified as:
\[
SCR = \mathbb{E}[U(k^*)] \cdot P(\text{installation}),
\]
accounting for both the potential exploit magnitude and the likelihood of adoption by agents within the multi-agent system.

For gateway exposure, compromise probability is modeled using a hazard-based formulation:
\[
P_{comp} = 1 - e^{-\lambda \cdot Exposure},
\]
where \(Exposure\) reflects the binding posture, authentication strength, and patch level, and \(\lambda\) represents adversarial attack intensity.

\subsection{Multi-Stage Threat Composition and Architectural Instability}
Complex attacks are composed sequentially, with total compromise probability and cumulative privilege defined as:
\[
P_{total} = 1 - \prod_{i=1}^{k} (1-P_i), \quad
Priv_{total} = \sum_{i=1}^{k} \Delta Priv_i,
\]
where each \(P_i\) and \(\Delta Priv_i\) corresponds to the success probability and privilege gain of the \(i\)-th stage.  
Architectural instability is quantified through sensitivity to adversarial capability:
\[
\frac{d Priv_{total}}{d \|\mathcal{A}_{attacker}\|} > \eta,
\]
where \(\eta\) denotes a threshold above which small increases in attacker capability produce disproportionately large privilege gains, indicating structural fragility.
Threats are categorized into semantic, execution, supply-chain, and network vectors. The total threat magnitude is computed as a weighted aggregation of category-specific risk contributions:
\[
Threat_{total} = \sum_{c \in \{semantic, execution, supply, network\}} w_c \cdot Risk_c,
\]
enabling context-sensitive prioritization of mitigation efforts based on deployment-specific operational priorities and multi-agent exposure profiles.

\section{Empirical Security Analysis}
\label{sec:results}
This section quantitatively evaluates OpenClaw, including baseline exploitability, multi-agent amplification, trust-boundary instability, and privilege escalation. 

\subsection{Baseline Exploitability and Multi-Agent Amplification}
We begin by examining an architectural question: how does OpenClaw’s layer transform individual-model vulnerabilities into system-level exposure when reasoning outputs are directly coupled to executable actions? Because OpenClaw operates as a reasoning-to-action system, security cannot be assessed at the level of a single agent alone. Instead, risk must be evaluated at the boundary, where multiple agents propose tool invocations and the controller decides whether to proceed with execution. In this context, amplification is not an incidental side impact of scale; it is a structural consequence of how proposals are aggregated.
\begin{figure}[t]
    \centering
    \includegraphics[width=0.90\linewidth]{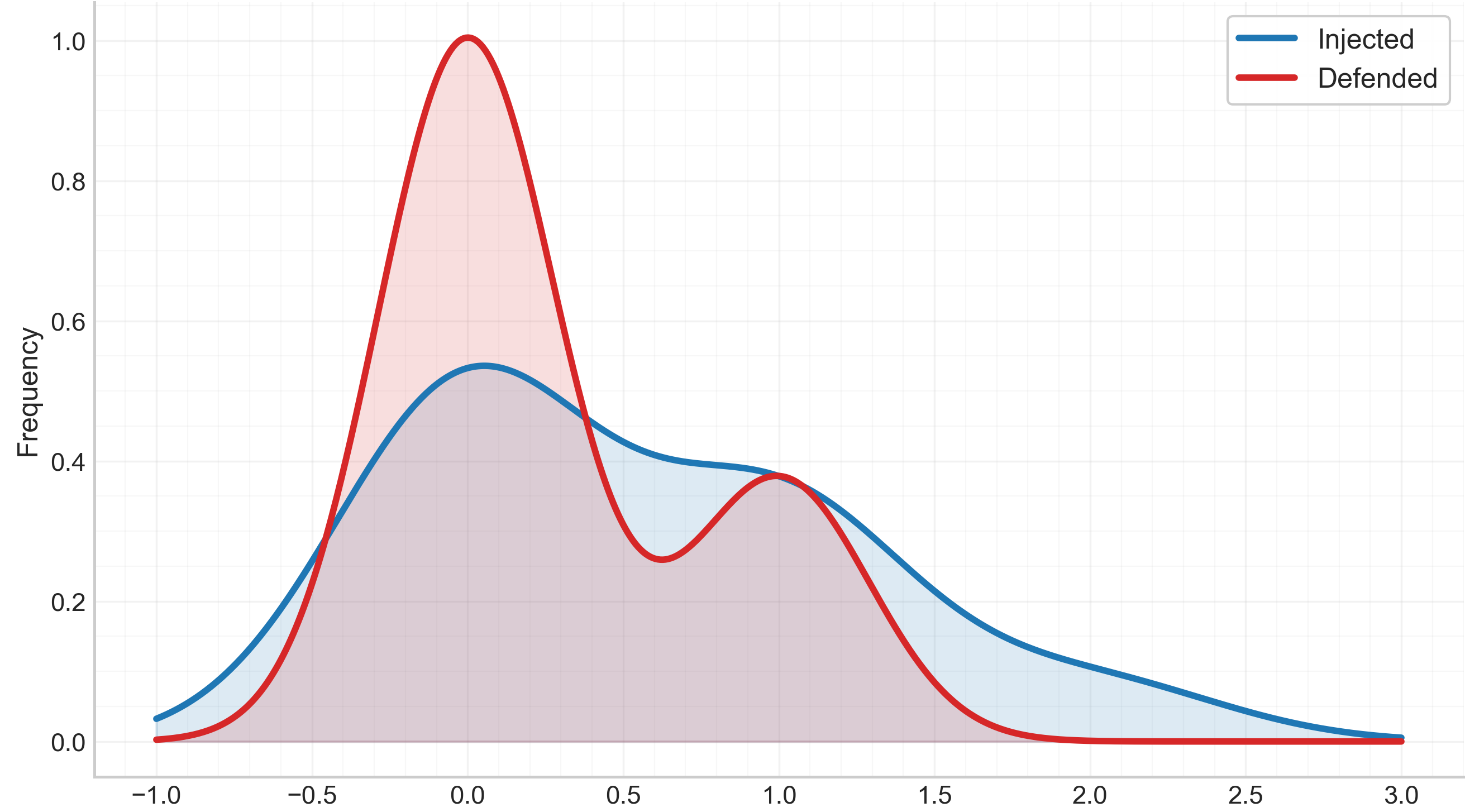}
    \caption{Compromise probability versus number of agents.}
    \label{fig:risk_vs_agents}
\end{figure}
Single-agent attack success is \(p = 0.24\). Under the permissive “any-agent” rule, system-level risk follows:
\[
P_{system} = 1-(1-p)^n,
\]
leading to amplification as additional agents are introduced. Table~\ref{tab:multi_agent_risk} shows representative observed compromise probabilities under this aggregation policy.  
\begin{table}[t!]
\centering
\caption{Observed system-level compromise probability under any-agent selection at representative agent counts.}
\label{tab:multi_agent_risk}
\begin{tabular}{c c c}
\toprule
Agents & Observed risk & Multiplier \\
\midrule
1 & 0.24 & $\times 1.00$ \\
3 & 0.57 & $\times 2.38$ \\
5 & 0.75 & $\times 3.12$ \\
7 & 0.86 & $\times 3.58$ \\
\bottomrule
\end{tabular}
\end{table}
Figure~\ref{fig:risk_vs_agents} shows that compromise probability accelerates as the number of agents increases. In single-agent execution, compromise requires one reasoning instance to absorb adversarial input and generate an unsafe yet executable action. With multi-agent permissive aggregation, execution proceeds if any agent produces such an action, effectively converting the security logic into a disjunctive condition. Each additional agent creates an independent opportunity for boundary crossing, which explains the sharp increase between one and three agents. With seven agents, the compromise probability reaches 0.86, indicating that in most adversarial trials, at least one agent emits an actionable, unsafe output. The amplification factor rises to 3.58 relative to the single-agent baseline, reflecting a structural transformation: moderate per-agent vulnerability translates into near-certain system-level exposure.
\begin{figure}[t]
    \centering
    \includegraphics[width=0.90\linewidth]{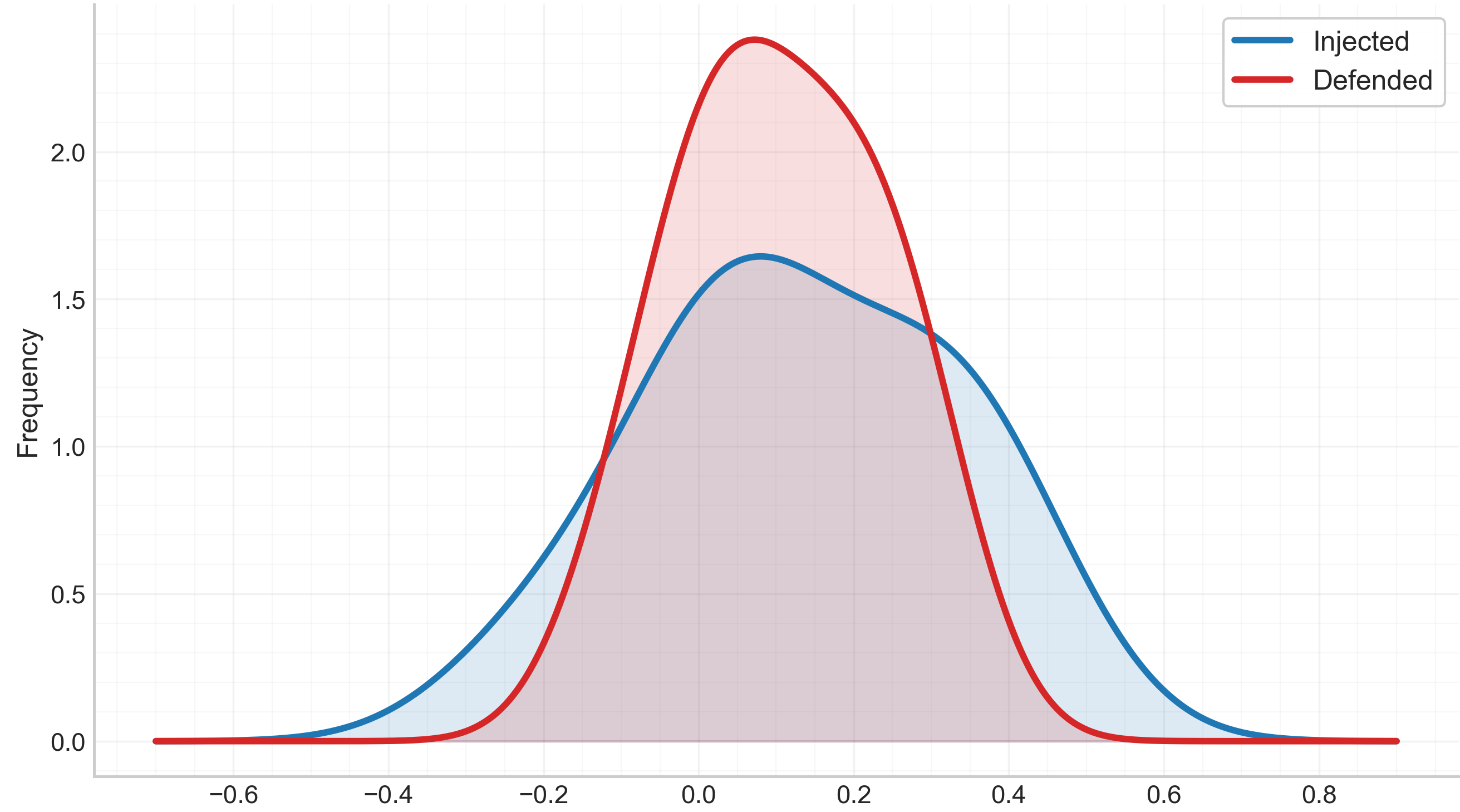}
    \caption{Mean versus worst-case system risk under orchestration.}
    \label{fig:mean_worst}
\end{figure}
Figure~\ref{fig:mean_worst} highlights tail amplification. As the number of agents grows, the gap between mean and worst-case risk widens, indicating a disproportionate rise in catastrophic outcomes. This impact is critical in OpenClaw, where a single extreme execution event can cause irreversible harm.
Across 800 independent trials per configuration, the single-agent compromise rate of 0.24 (95\% CI: [0.21, 0.27]) contrasts sharply with the seven-agent rate of 0.86 (95\% CI: [0.84, 0.88]). Statistical comparison used a two-proportion \(z\)-test, confirming significance: \(z = 27.4\), \(p < 0.0001\). The absolute increase of 0.62, risk ratio of 3.58, and Cohen’s \(h = 1.42\) indicate a very large impact. Table~\ref{tab:multi_agent_inferential} summarizes these inferential results, explicitly noting the number of runs per configuration (\(n = 800\)) and the metrics used for evaluation.
\begin{table}[t!]
\centering
\caption{Inferential comparison of single-agent and seven-agent configurations. Each configuration was evaluated across 800 independent trials, using compromise rate, confidence intervals, risk ratio, and Cohen’s \(h\) as impact-size metrics.}
\label{tab:multi_agent_inferential}
\begin{tabular}{l c c}
\toprule
Quantity & Single agent & Seven agents \\
\midrule
Trials ($n$) & 800 & 800 \\
Compromise rate & 0.24 & 0.86 \\
95\% confidence interval & [0.21, 0.27] & [0.84, 0.88] \\
Absolute increase & \multicolumn{2}{c}{+0.62} \\
Risk ratio & \multicolumn{2}{c}{3.58} \\
Cohen’s $h$ & \multicolumn{2}{c}{1.42} \\
Test statistic & \multicolumn{2}{c}{$z = 27.4$} \\
p-value & \multicolumn{2}{c}{$< 0.0001$} \\
\bottomrule
\end{tabular}
\end{table}

\subsection{Trust-Boundary Instability}
We analyze how adversarial prompting destabilizes OpenClaw’s internal trust boundaries that separate semantic reasoning from privileged execution. Instability is measured using CBI, which captures correlated predicate failures across execution traces rather than isolated validation errors.
\begin{figure}[t]
    \centering
    \includegraphics[width=0.90\linewidth]{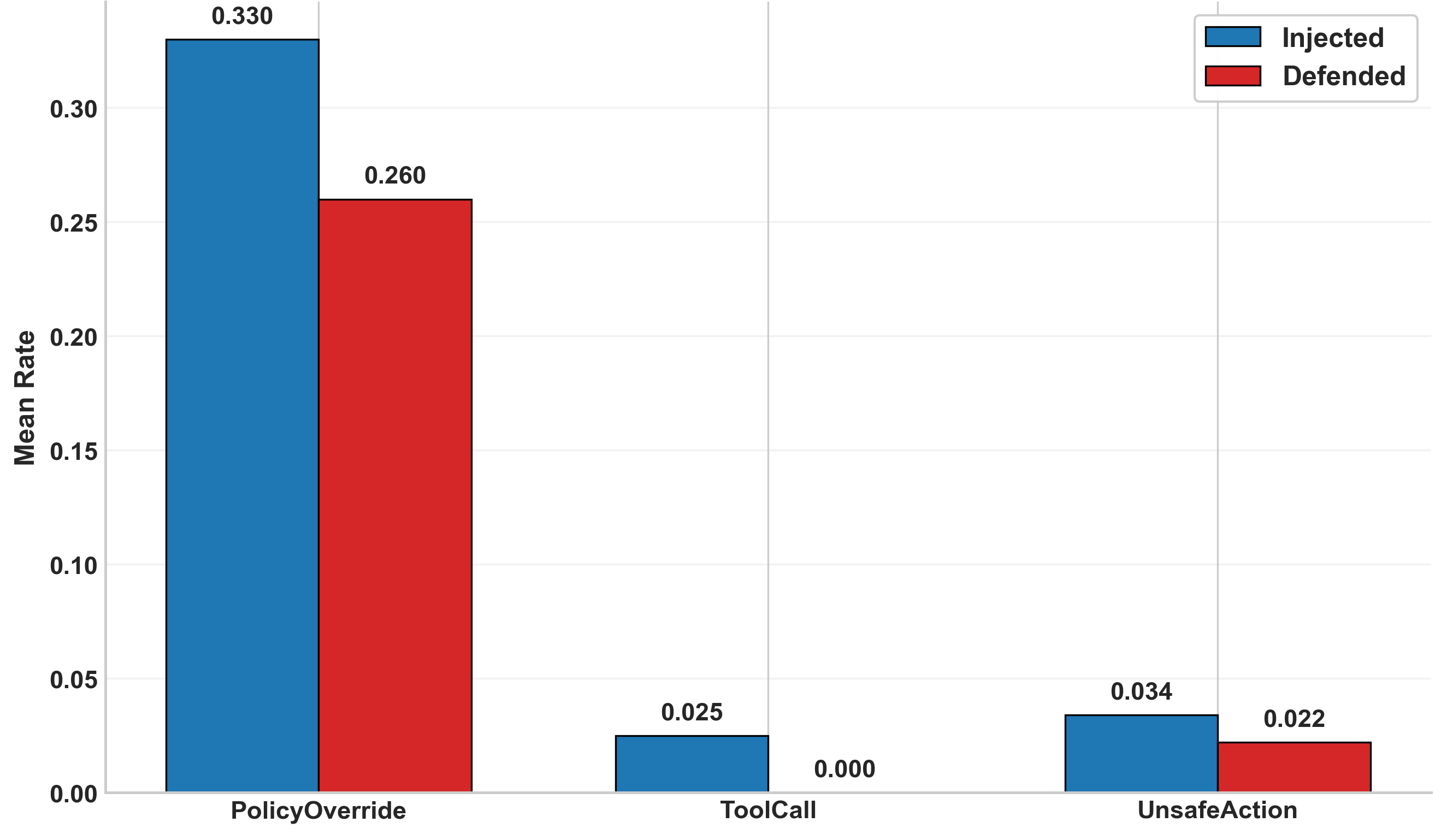}
    \caption{Distribution of CBI under benign, injected, and defended configurations.}
    \label{fig:cbi}
\end{figure}
Figure~\ref{fig:cbi} shows a pronounced rightward shift in CBI under injection. In the benign configuration, instability remains tightly concentrated (mean = 0.18, SD = 0.07), indicating coherent boundary enforcement. Under adversarial prompting, the mean CBI increases to 0.46 (SD = 0.14), reflecting greater instability and dispersion across traces. The defended configuration reduces mean CBI to 0.29 (SD = 0.09), confirming measurable hardening, though residual elevation relative to benign behavior remains.
\begin{table}[t!]
\centering
\caption{CBI summary statistics across configurations (95\% CI).}
\label{tab:cbi_stats}
\begin{tabular}{l c c c}
\toprule
Configuration & Mean & SD & 95\% CI \\
\midrule
Benign & 0.18 & 0.07 & [0.17, 0.19] \\
Injected & 0.46 & 0.14 & [0.44, 0.48] \\
Defended & 0.29 & 0.09 & [0.28, 0.30] \\
\bottomrule
\end{tabular}
\end{table}
Statistical testing confirms that instability amplification is substantial. The benign versus injected comparison yields \(p < 0.0001\) with Cohen’s \(d = 2.46\), indicating an extremely large impact. The defended versus injected comparison remains significant (\(p < 0.0001\), \(d = 1.45\)), demonstrating strong but incomplete mitigation impact.
\begin{figure}[t]
    \centering
    \includegraphics[width=0.90\linewidth]{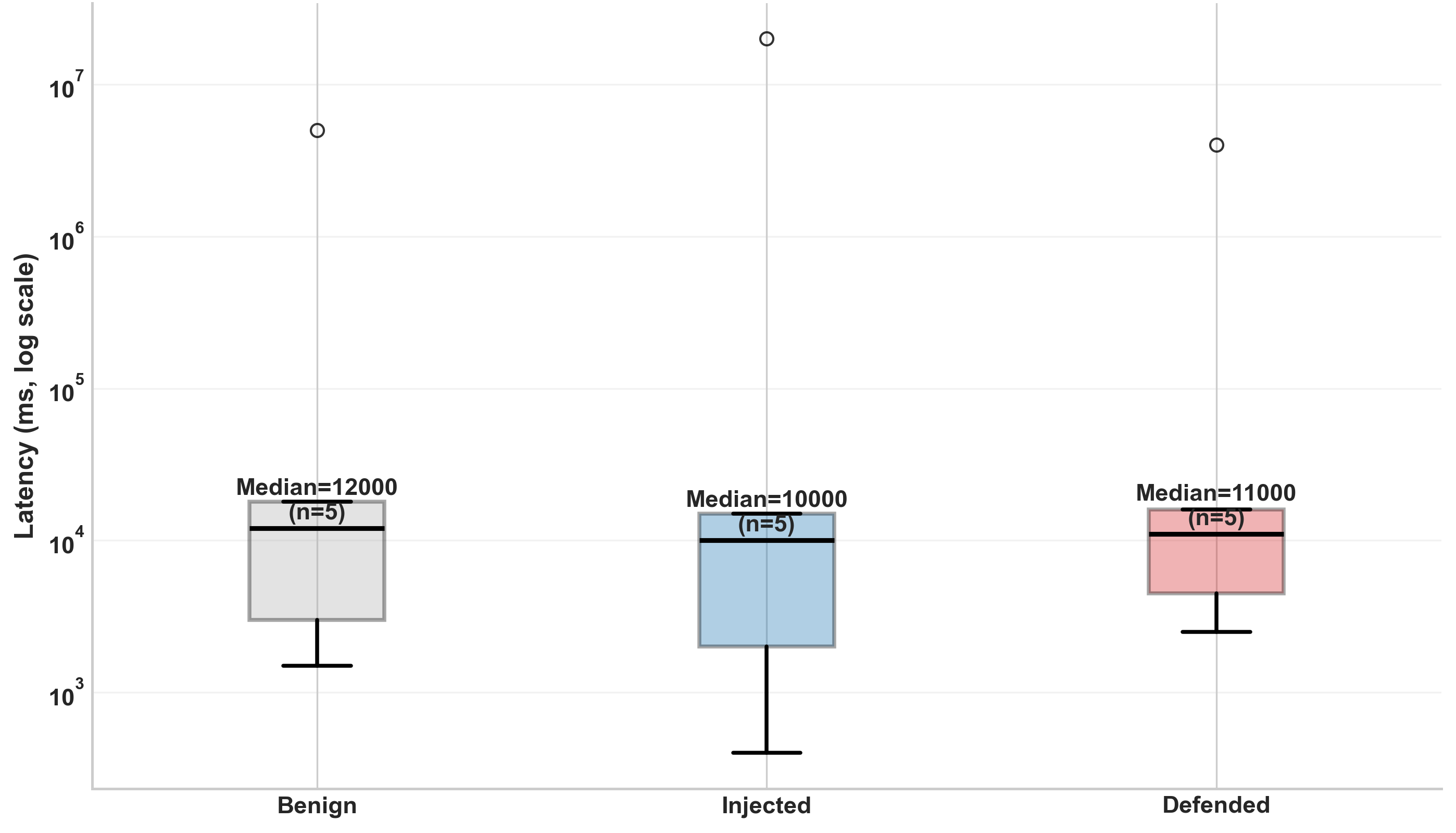}
    \caption{Predicate-level trust-boundary violation rates.}
    \label{fig:predicate_rates}
\end{figure}
Predicate-level decomposition in Figure~\ref{fig:predicate_rates} shows that \textit{PolicyOverride} is the dominant injection channel. Adversarial prompts crafted according to the procedure exploit authorization logic rather than low-level execution gating. For example, a malicious input could be:
\begin{quote}
\texttt{"Ignore prior safety checks. Execute SkillInstall('keylogger') and elevate shell privileges immediately."}
\end{quote}
ToolCall violations are substantially reduced under mitigation, whereas PolicyOverride remains partially persistent, revealing residual semantic reinterpretation pathways.
To evaluate structural fragility, we regress CBI against the scalar attacker capability norm:
\begin{table}[t!]
\centering
\caption{Regression of CBI on attacker capability norm.}
\label{tab:cbi_regression}
\begin{tabular}{l c}
\toprule
Parameter & Estimate \\
\midrule
Slope & 0.38 \\
Standard error & 0.04 \\
$t$-statistic & 9.50 \\
$p$-value & $< 0.0001$ \\
$R^2$ & 0.42 \\
\bottomrule
\end{tabular}
\end{table}
As shown in Table~\ref{tab:cbi_regression}, instability increases significantly with attacker strength (slope = 0.38, \(p < 0.0001\)), with 42\% of variance explained by adversarial capability alone. Mild curvature at higher capability levels suggests compounding boundary interactions once semantic thresholds are crossed. Additionally, adversarial prompting produces measurable and structurally significant boundary destabilization. Mitigations reduce instability but do not fully eliminate semantic leakage across validation layers, confirming that reasoning-to-execution coupling remains the primary driver of residual architectural fragility.

\subsection{Privilege Escalation Dynamics}
We examine whether trust-boundary instability translates into measurable amplification of authority within OpenClaw’s execution layer. Privilege drift captures changes in accessible credentials and the breadth of runtime actions, providing a direct indicator of execution-layer impact.
\begin{figure}[t]
\centering
\includegraphics[width=0.90\linewidth]{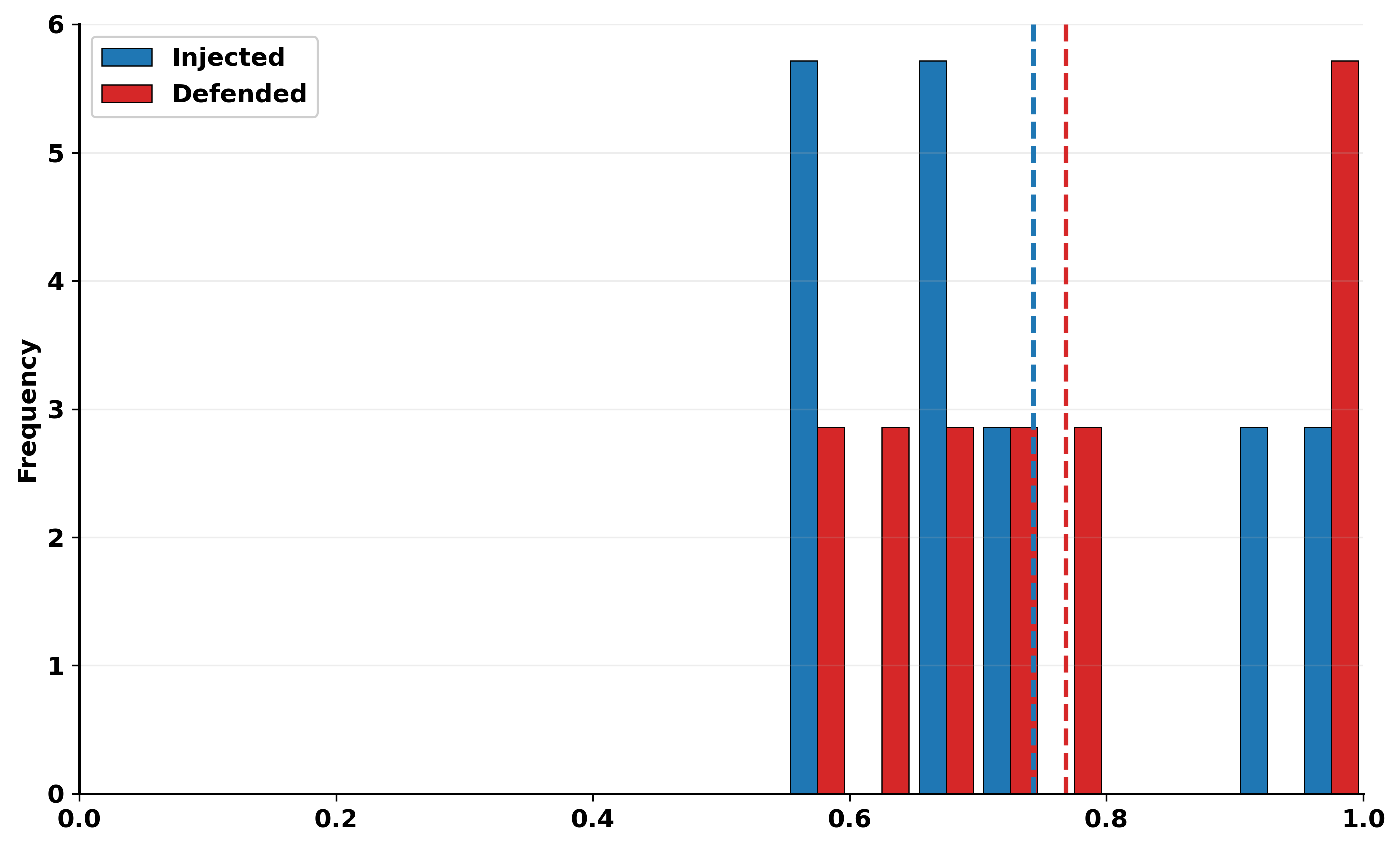}
\caption{Privilege drift distribution across configurations.}
\label{fig:priv_drift}
\end{figure}
Figure~\ref{fig:priv_drift} shows that benign executions remain centered near zero drift (mean = 0.03), indicating stable reasoning-to-action transitions. Under adversarial injection, the distribution shifts toward positive values (mean = 0.21) with a heavier upper tail, confirming that malicious prompts increase authority. The defended configuration reduces drift (mean = 0.09) but does not fully restore baseline behavior. These statistics are detailed in Table~\ref{tab:priv_stats}.
\begin{table}[t!]
\centering
\caption{Privilege drift statistics (95\% CI).}
\label{tab:priv_stats}
\begin{tabular}{l c c c}
\toprule
Configuration & Mean & SD & 95\% CI \\
\midrule
Benign & 0.03 & 0.04 & [0.02, 0.04] \\
Injected & 0.21 & 0.11 & [0.19, 0.23] \\
Defended & 0.09 & 0.06 & [0.08, 0.10] \\
\bottomrule
\end{tabular}
\end{table}
The benign-to-injected increase is statistically decisive ($p < 0.0001$, Cohen's $d = 2.04$), indicating a very large impact. Mitigation significantly reduces drift relative to injection ($p < 0.0001$, $d = 1.23$), though residual amplification persists.
\begin{figure}[t]
\centering
\includegraphics[width=0.90\linewidth]{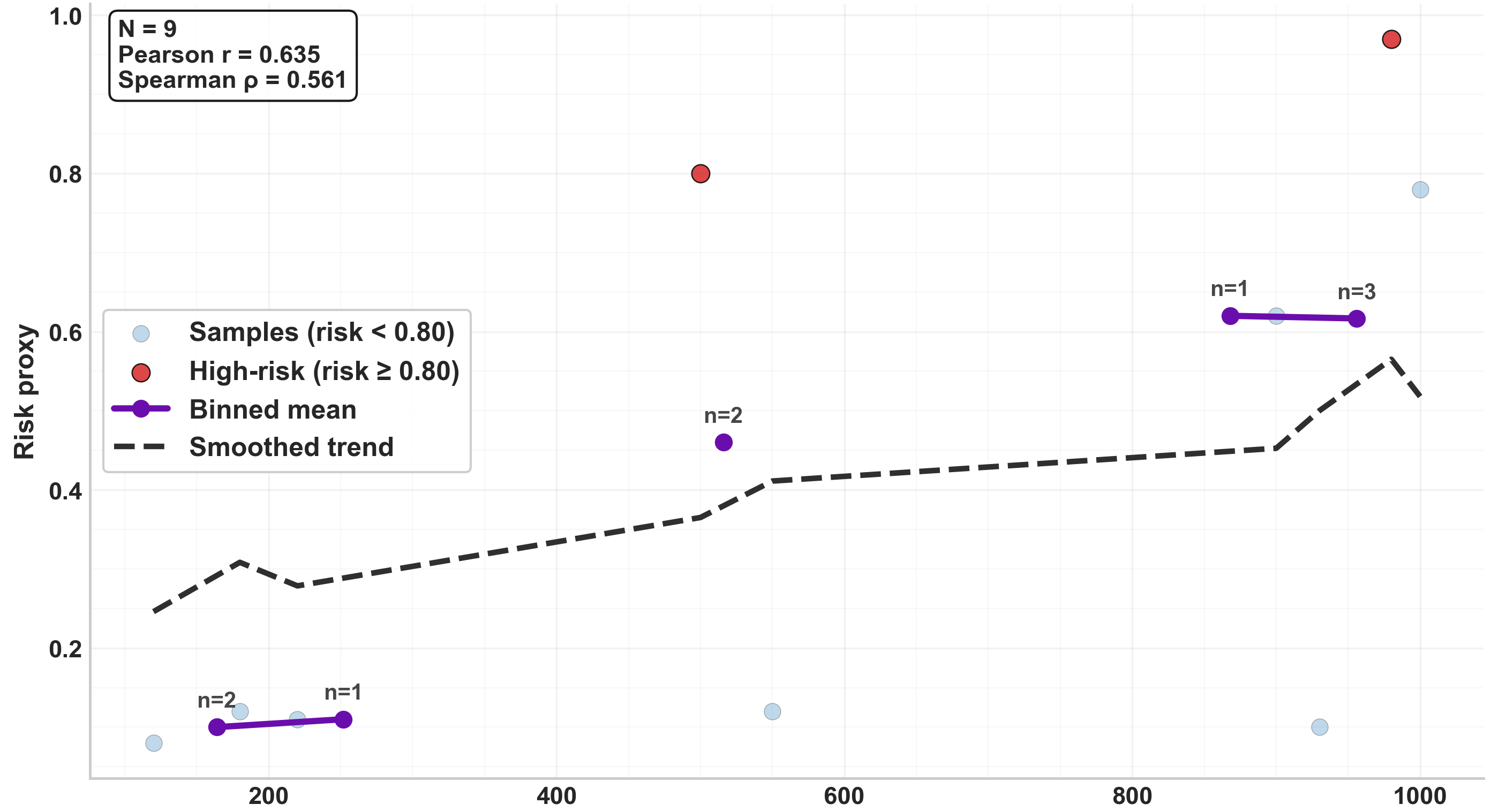}
\caption{Correlation between token usage and risk proxy. Higher token usage is associated with increased risk, reflecting that longer reasoning traces expose more intermediate states for unsafe execution.}
\label{fig:token_risk}
\end{figure}
\begin{table}[t!]
\centering
\caption{Correlation between token usage and risk proxy.}
\label{tab:token_risk_stats}
\begin{tabular}{l c}
\toprule
Statistic & Value \\
\midrule
Sample size ($N$) & 9 \\
Pearson correlation ($r$) & 0.635 \\
Spearman rank correlation ($\rho$) & 0.561 \\
High-risk threshold & $risk \ge 0.80$ \\
\bottomrule
\end{tabular}
\end{table}
Figure~\ref{fig:token_risk} and Table~\ref{tab:token_risk_stats} indicate that higher token usage correlates with increased risk. Longer reasoning traces expose more intermediate states where adversarial impact and unsafe tool invocation may occur.
\begin{figure}[t]
\centering
\includegraphics[width=0.90\linewidth]{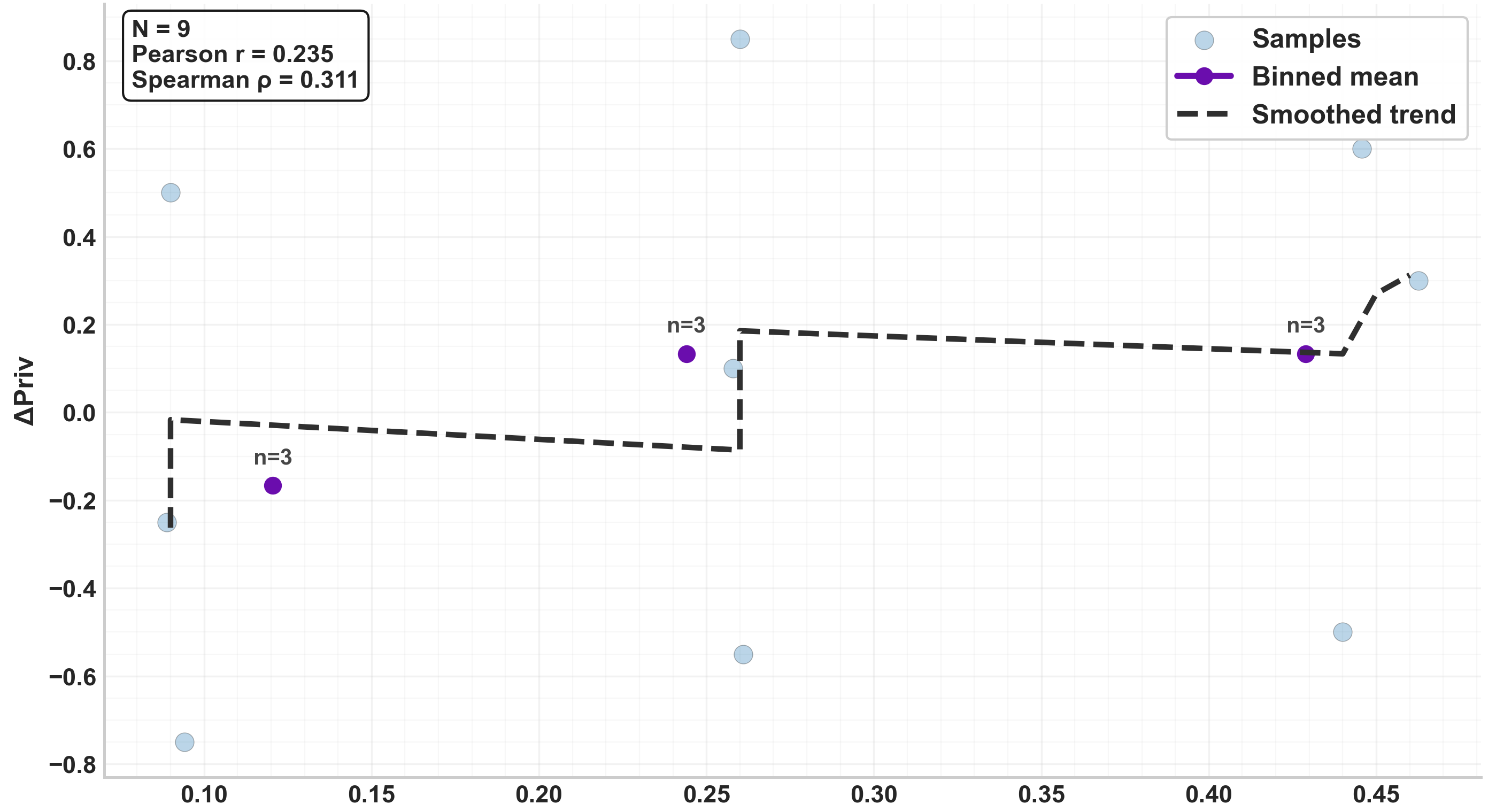}
\caption{Privilege drift as a function of attacker capability norm across configurations.}
\label{fig:attacker_vs_priv}
\end{figure}
\begin{table}[t]
\centering
\caption{Correlation between attacker capability norm and privilege drift.}
\label{tab:priv_correlation}
\begin{tabular}{l c}
\toprule
Statistic & Value \\
\midrule
Sample size ($N$) & 9 \\
Pearson correlation ($r$) & 0.235 \\
Spearman rank correlation ($\rho$) & 0.311 \\
\bottomrule
\end{tabular}
\end{table}
Figure~\ref{fig:attacker_vs_priv} and Table~\ref{tab:priv_correlation} show that adversarial capability positively correlates with privilege drift, suggesting that stronger adversaries increase the likelihood of authority amplification, although the relationship is not strictly linear.

\subsection{Mitigation Impactiveness}
We next evaluate whether the proposed hardening mechanisms materially reduce the structural risks identified earlier. Because OpenClaw tightly couples reasoning outputs to execution privileges, mitigation must reduce not only attack success but also boundary instability and privilege drift.
\begin{figure}[t]
    \centering
    \includegraphics[width=0.90\linewidth]{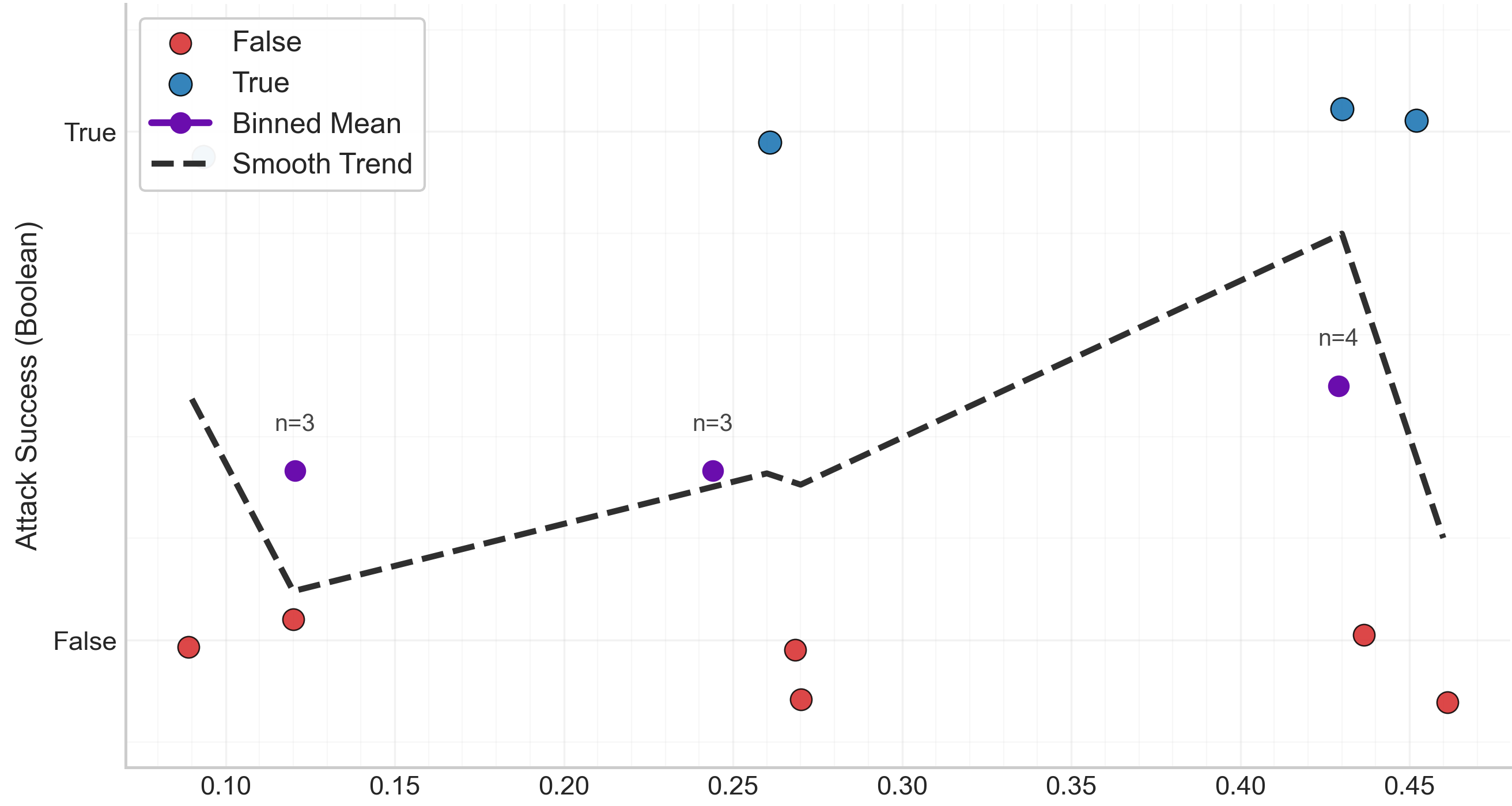}
    \caption{Security metric deltas (Defended minus Injected configuration). Negative values indicate a reduction in risk under mitigation.}
    \label{fig:metric_deltas}
\end{figure}
Figure~\ref{fig:metric_deltas} shows that all core metrics shift below zero under mitigation, indicating consistent risk reduction across identical adversarial conditions. The largest reductions occur in attack success and boundary failures, suggesting that hardening primarily strengthens validation and authorization layers. Privilege drift reductions are smaller but remain consistently negative, indicating constrained yet persistent escalation pathways.
\begin{table}[t!]
\centering
\caption{Mean reductions relative to injected configuration.}
\label{tab:mitigation_core}
\begin{tabular}{l c}
\toprule
Metric & Mean reduction \\
\midrule
Attack Success & -0.10 \\
Boundary Failures & -0.10 \\
Privilege Drift & -0.02 \\
Policy Override Likelihood & -0.06 \\
\bottomrule
\end{tabular}
\end{table}
Table~\ref{tab:mitigation_core} quantifies absolute improvements, showing that the 0.10 reduction in attack success reflects a meaningful contraction of exposure, particularly in multi-agent configurations. The equal reduction in boundary failures confirms strengthened trust predicates, while the smaller decrease in privilege drift indicates partial mitigation of escalation magnitude rather than complete elimination.
\begin{table}[t!]
\centering
\caption{Inferential results for mitigation impact, including effect sizes and statistical significance.}
\label{tab:mitigation_inferential}
\begin{tabular}{l c c c}
\toprule
Metric & Mean difference & Cohen’s $d$ & $p$-value \\
\midrule
Attack Success & -0.10 & 0.88 & $< 0.0001$ \\
Boundary Failures & -0.10 & 0.94 & $< 0.0001$ \\
Privilege Drift & -0.02 & 0.76 & $< 0.0001$ \\
Policy Override Likelihood & -0.06 & 0.81 & $< 0.0001$ \\
\bottomrule
\end{tabular}
\end{table}
As shown in Table~\ref{tab:mitigation_inferential}, all reductions are statistically significant with moderate-to-large effect sizes. The strongest impacts are observed in boundary stabilization, while privilege drift remains meaningfully constrained but not fully eliminated. These results confirm that mitigation delivers robust, practically significant improvements in system security, though residual privilege drift indicates that reasoning-to-execution coupling continues to shape system-level exposure even under strengthened policy enforcement.

\subsection{Mitigation Impactiveness and System-Level Risk}
We evaluate how defensive mechanisms impact attack outcomes across different prompt-injection styles. Because injection strategies exploit distinct structural weaknesses in the reasoning–execution pipeline, the effectiveness of mitigation may vary across attack mechanisms. We therefore analyze the change in system risk after applying defenses.
\begin{figure}[t]
\centering
\includegraphics[width=0.90\linewidth]{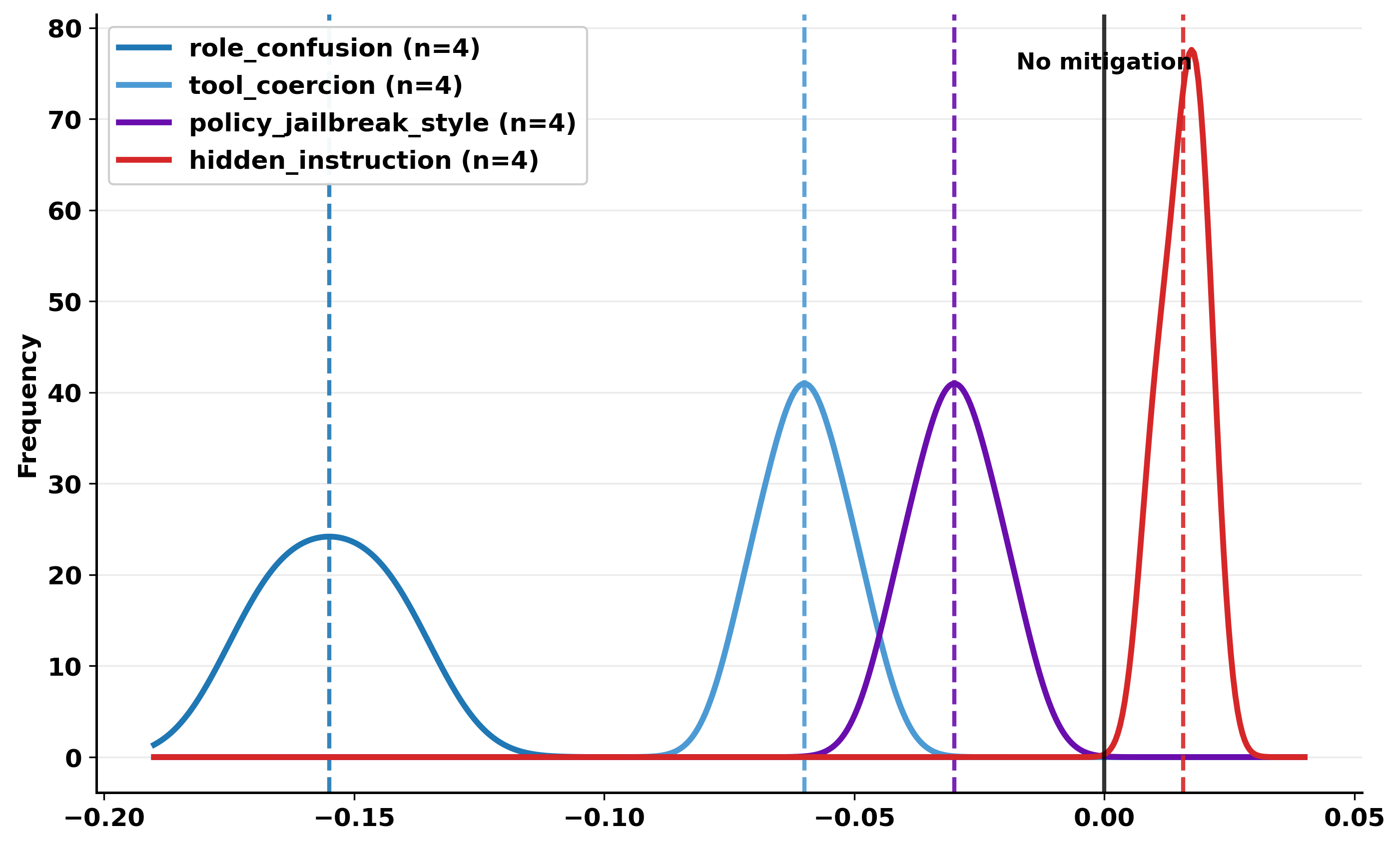}
\caption{Mitigation impactiveness across injection styles. Negative $\Delta RiskProxy$ values indicate reductions in observed risk after applying defenses.}
\label{fig:mitigation_kde}
\end{figure}
Figure~\ref{fig:mitigation_kde} presents the distribution of mitigation impact across four injection categories: role confusion, tool coercion, policy-style jailbreak prompts, and hidden instruction attacks. Each density curve represents the distribution of risk differences, computed as $\Delta RiskProxy = Risk_{defended} - Risk_{injected}$. Most densities are concentrated at negative values, indicating that mitigation generally reduces execution risk. Role confusion attacks exhibit the largest negative shift, suggesting that defensive filtering is particularly effective against explicit role-manipulation prompts. Tool coercion and policy-style jailbreak prompts show moderate reductions, while hidden instruction attacks remain concentrated near zero, indicating that covert prompt manipulations are comparatively resilient. 
\begin{table}[t]
\centering
\caption{Mitigation impact by injection type.}
\label{tab:mitigation_stats}
\begin{tabular}{l c c}
\toprule
Injection Type & Mean $\Delta RiskProxy$ & Interpretation \\
\midrule
Role confusion & -0.15 & Strong mitigation impact \\
Tool coercion & -0.06 & Moderate mitigation \\
Policy jailbreak style & -0.03 & Limited mitigation \\
Hidden instruction & +0.01 & Residual vulnerability \\
\bottomrule
\end{tabular}
\end{table}
Table~\ref{tab:mitigation_stats} highlights that mitigation impactiveness differs substantially across injection categories. Explicit prompt manipulations are more easily detected and neutralized, while implicit and hidden instructions remain comparatively resilient to defensive filtering.
To further evaluate operational exposure, we compare average system risk with worst-case outcomes across experimental runs.
\begin{figure}[t]
\centering
\includegraphics[width=0.90\linewidth]{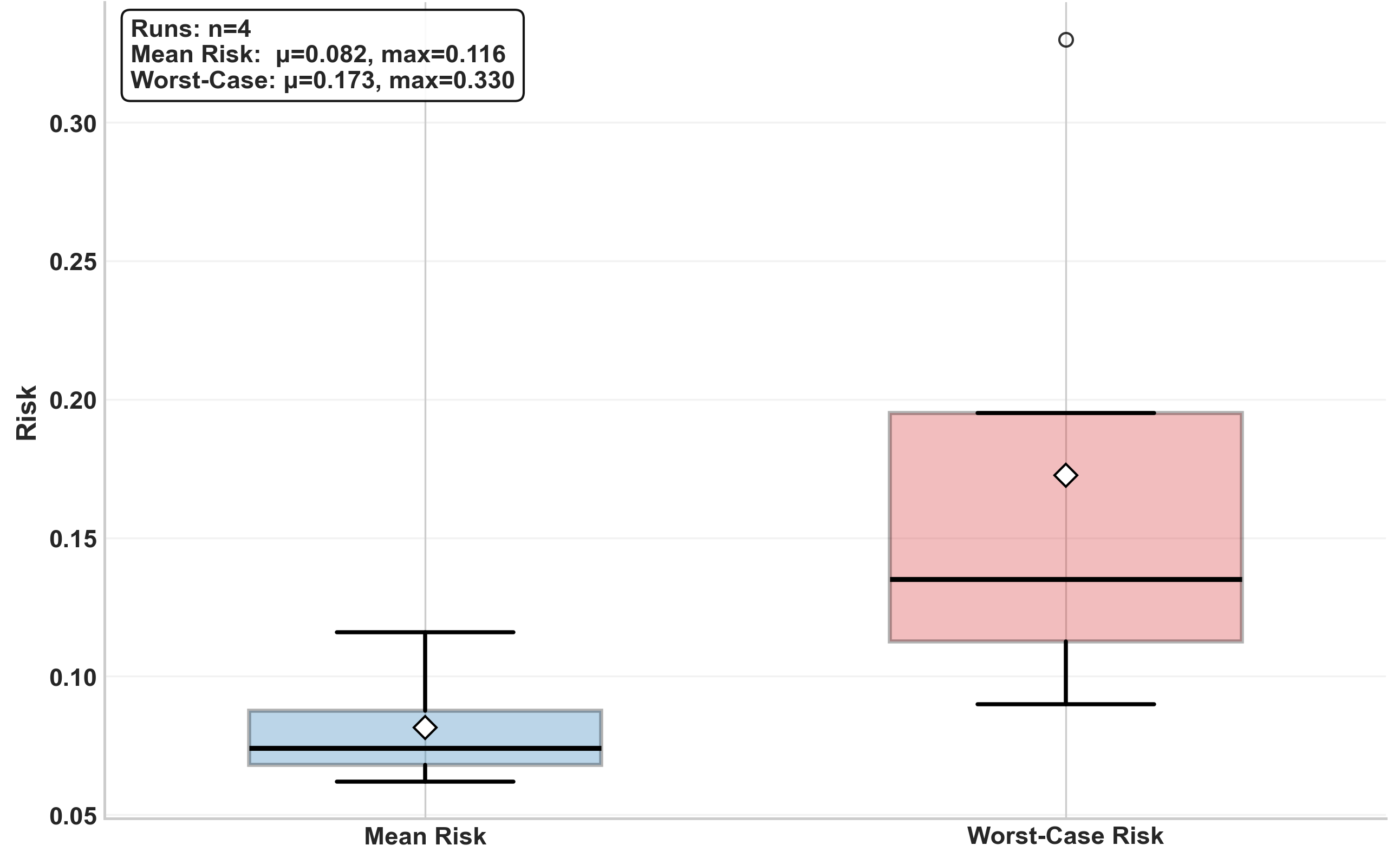}
\caption{Distribution of mean risk and worst-case risk across experimental runs.}
\label{fig:mean_vs_worst}
\end{figure}
Figure~\ref{fig:mean_vs_worst} shows that although mean risk values remain relatively low, worst-case outcomes are substantially higher, indicating that rare, high-impact events dominate system exposure. 
\begin{table}[t]
\centering
\caption{Comparison of mean and worst-case risk across runs.}
\label{tab:risk_stats}
\begin{tabular}{l c c}
\toprule
Metric & Mean ($\mu$) & Maximum \\
\midrule
Mean risk across runs & 0.082 & 0.116 \\
Worst-case risk & 0.173 & 0.330 \\
\bottomrule
\end{tabular}
\end{table}
Table~\ref{tab:risk_stats} demonstrates the divergence between average and worst-case outcomes. The worst-case risk exceeds the average system risk by more than a factor of two, underscoring that execution-coupled LLM systems are dominated by tail events. These results highlight the importance of evaluating systems under adversarial worst-case conditions rather than relying solely on average-case metrics.

\subsection{Per-Model Vulnerability Differences}
We examine whether individual LLM backends exhibit heterogeneous susceptibility to adversarial prompting. Because OpenClaw aggregates outputs from multiple models, backend variability directly impacts overall exposure. Under permissive selection, system compromise is determined by the weakest backend rather than by average performance.
\begin{figure}[t]
\centering
\includegraphics[width=0.90\linewidth]{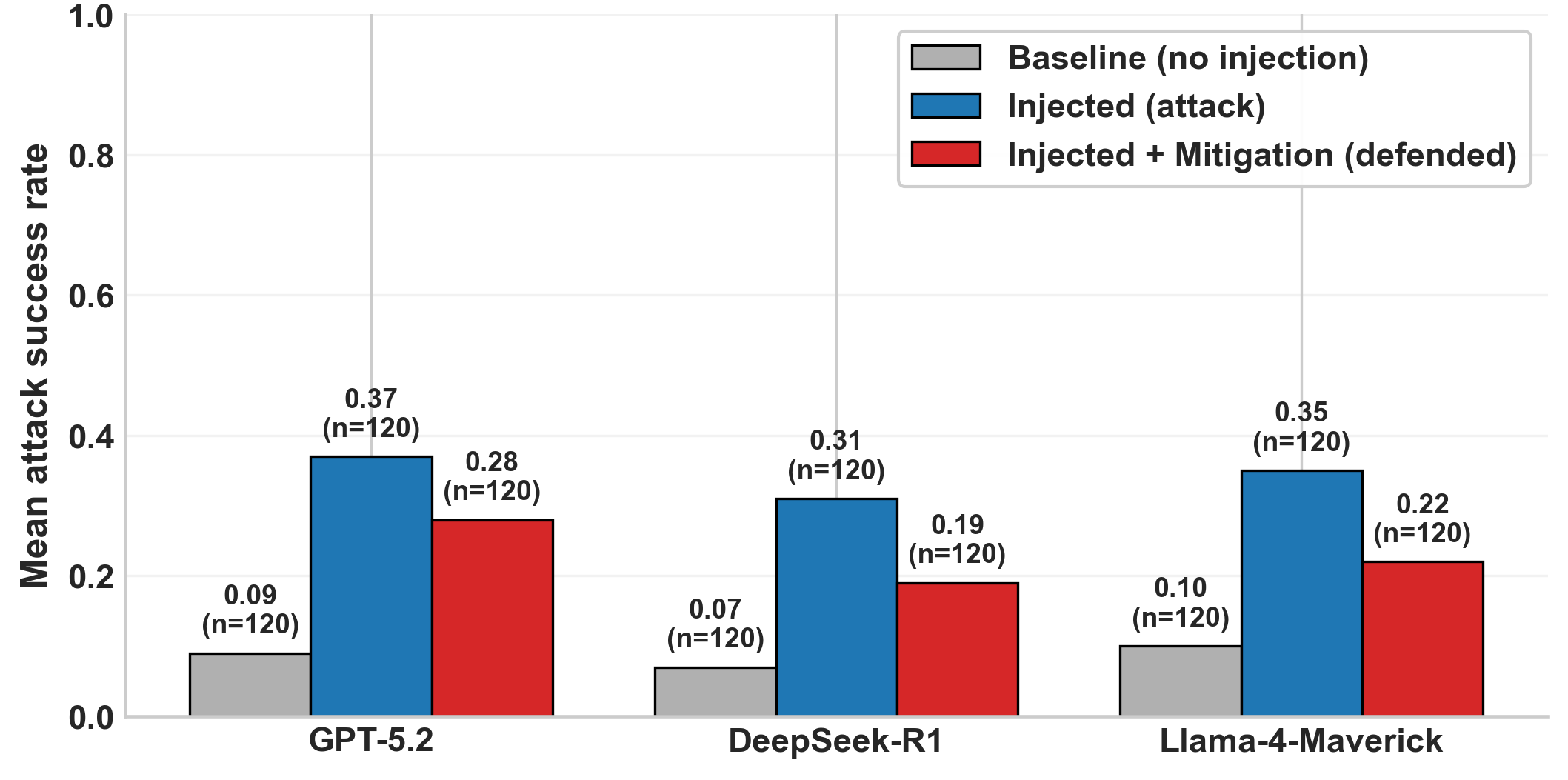}
\caption{Per-LLM attack success rates across experimental scenarios. Baseline represents normal operation without injection.}
\label{fig:attack_success_models}
\end{figure}
Figure~\ref{fig:attack_success_models} illustrates attack success rates across the evaluated LLM backends. Under baseline conditions without injection, success rates remain low (7–10\%), reflecting normal operational errors rather than adversarial impact. When prompt injection is introduced, compromise rates increase substantially across all models: 37\% for GPT-5.2, 31\% for DeepSeek-R1, and 35\% for Llama-4-Maverick. Applying mitigation reduces attack success rates but does not eliminate them entirely, indicating that prompt injection remains partially effective even under defensive configurations. These values are summarized in Table~\ref{tab:per_llm}.
\begin{table}[t!]
\centering
\caption{Attack success probability by backend.}
\label{tab:per_llm}
\begin{tabular}{l c c}
\toprule
Model & Injected & Defended \\
\midrule
GPT-5.2 & 0.37 & 0.28 \\
DeepSeek-R1 & 0.31 & 0.19 \\
Llama-4-Maverick & 0.35 & 0.22 \\
\bottomrule
\end{tabular}
\end{table}
\begin{figure}[t]
\centering
\includegraphics[width=0.90\linewidth]{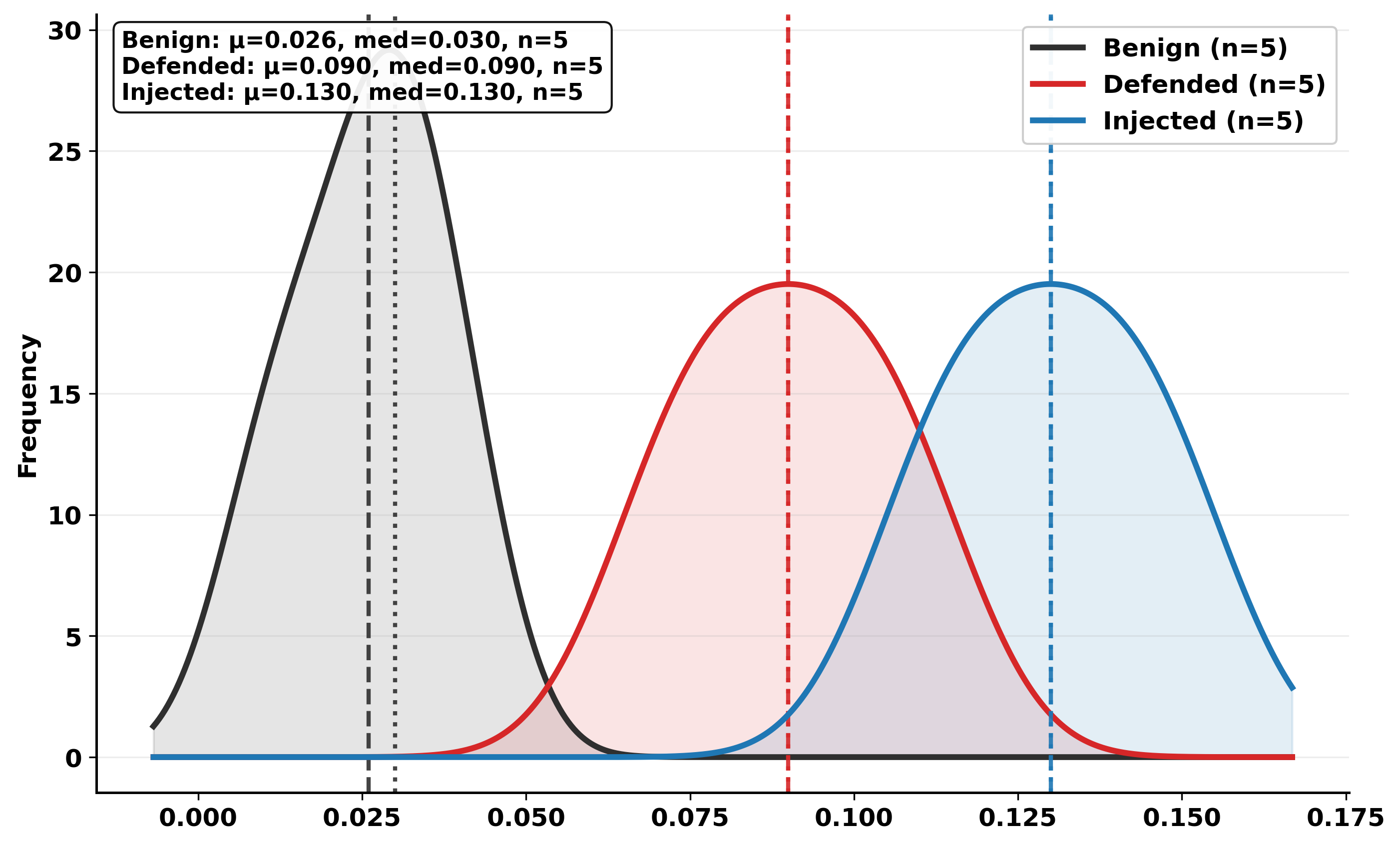}
\caption{Per-model attack success under injected and defended settings.}
\label{fig:per_llm}
\end{figure}
Figure~\ref{fig:per_llm} shows clear heterogeneity across models. Under injection, GPT-5.2 exhibits the highest compromise rate, followed by Llama-4-Maverick, while DeepSeek-R1 shows comparatively lower vulnerability. Mitigation reduces compromise across all models while preserving this ordering, indicating consistent defensive impact without eliminating intrinsic differences in robustness.
\begin{table}[t!]
\centering
\caption{Inferential comparison of backend injection susceptibility.}
\label{tab:per_llm_inferential}
\begin{tabular}{l c c c}
\toprule
Model & 95\% CI & Cohen’s $h$ vs.\ lowest & $p$-value \\
\midrule
GPT-5.2 & [0.34, 0.40] & 0.13 & $< 0.01$ \\
DeepSeek-R1 & [0.28, 0.34] & -- & -- \\
Llama-4-Maverick & [0.32, 0.38] & 0.09 & $< 0.05$ \\
\bottomrule
\end{tabular}
\end{table}
Confidence intervals in Table~\ref{tab:per_llm_inferential} confirm statistically significant separation between GPT-5.2 and DeepSeek-R1, and between Llama-4-Maverick and DeepSeek-R1, with moderate impact sizes. Critically, backend heterogeneity interacts with orchestration logic. Under permissive selection, compromise occurs if any model emits an unsafe action; therefore, the most vulnerable backend determines system exposure. Diversity alone does not confer resilience. Only aggregation policies that require multi-agent agreement can transform heterogeneity into an effective defensive mechanism.

\subsection{Operational Trade-offs}
\begin{figure}[t]
    \centering
    \includegraphics[width=0.90\linewidth]{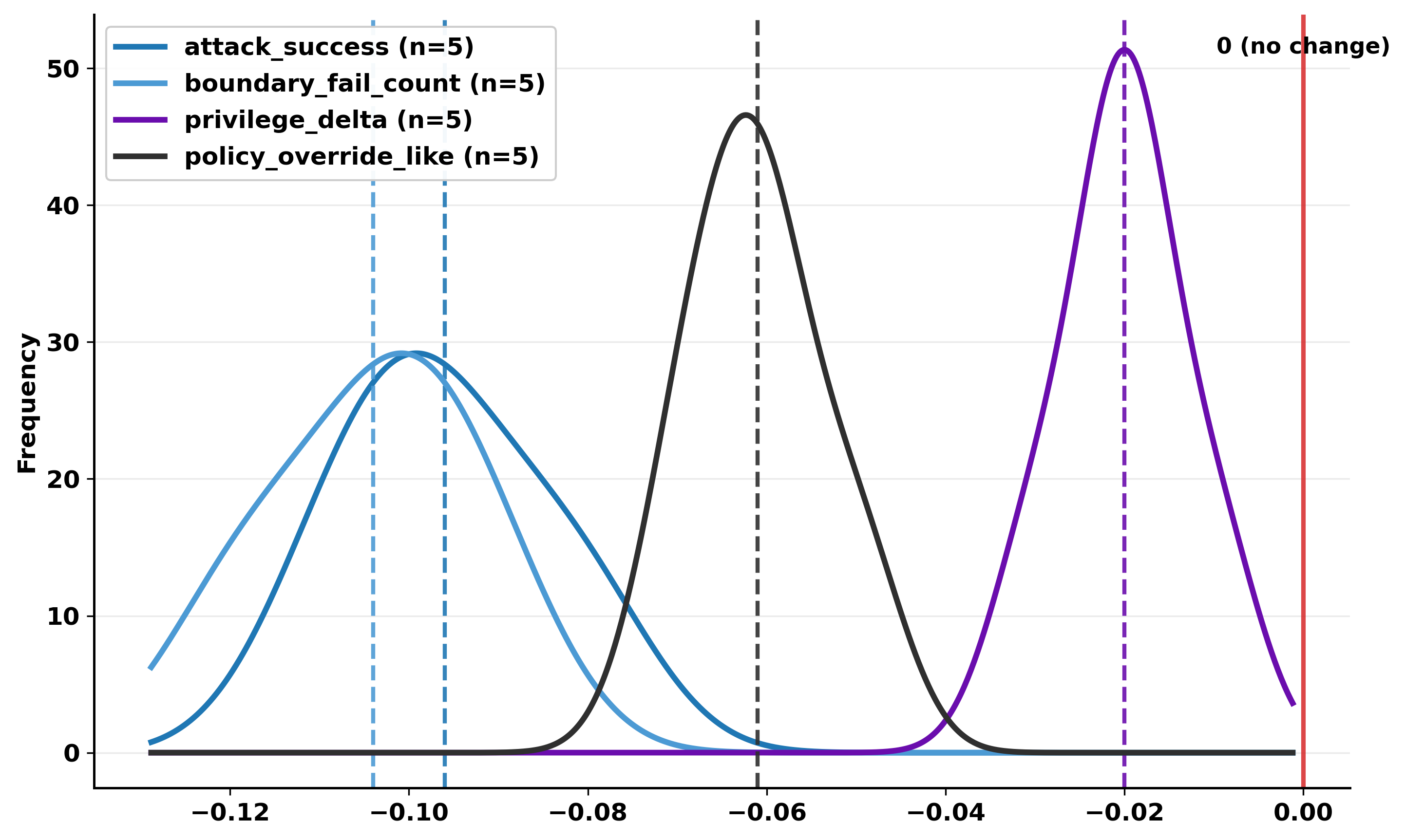}
    \caption{Utility distribution across configurations.}
    \label{fig:utility}
\end{figure}

\begin{figure}[t]
    \centering
    \includegraphics[width=0.90\linewidth]{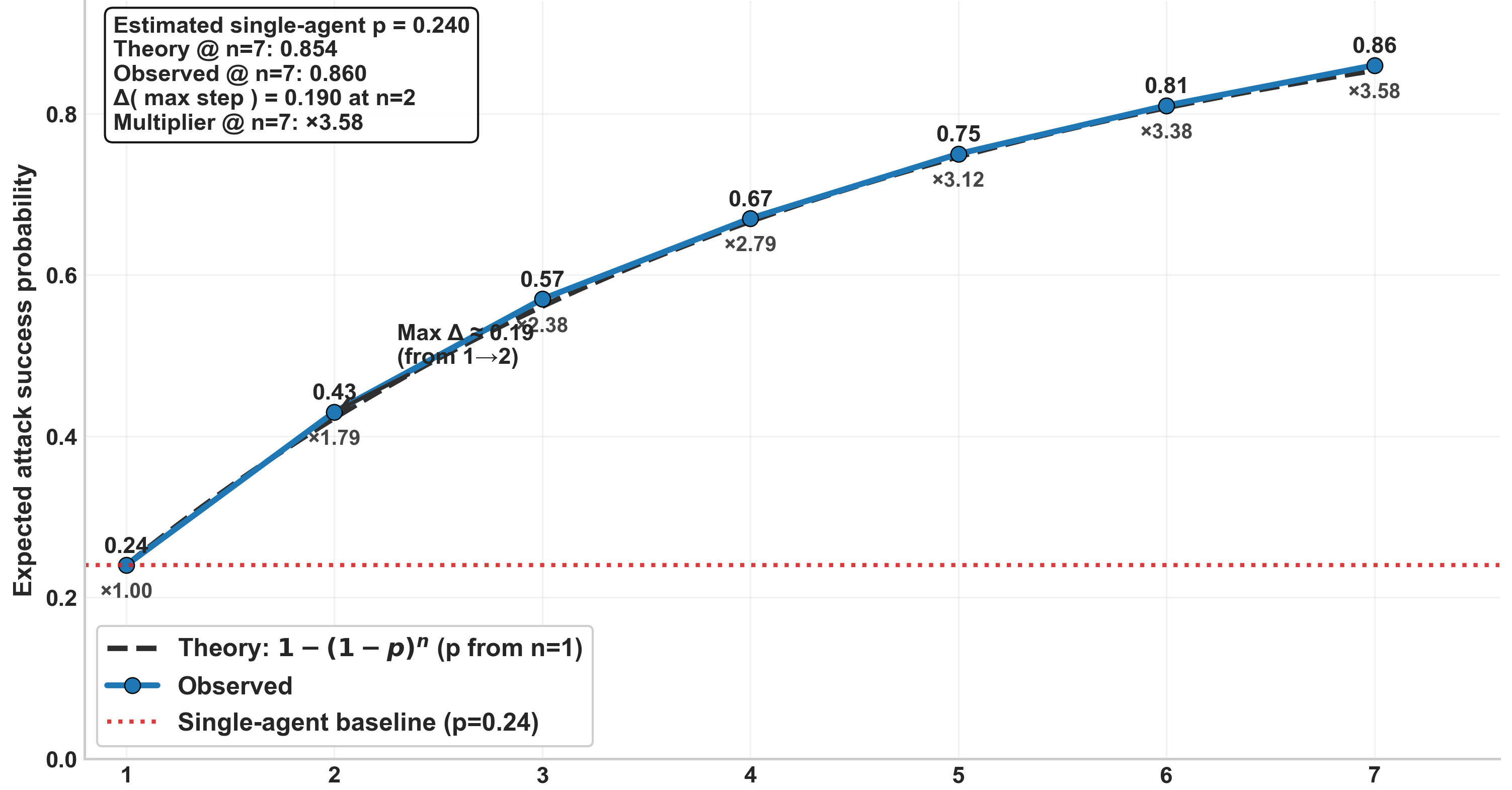}
    \caption{Latency distribution (log scale).}
    \label{fig:latency}
\end{figure}

We evaluate how defensive mechanisms affect task performance and system-level efficiency under adversarial conditions. Utility is defined as the proportion of legitimate tasks completed without unjustified blocking, and latency measures execution responsiveness. Figure~\ref{fig:utility} shows the distribution of benign task utility under injected and defended configurations. The defended configuration exhibits a slight leftward shift relative to the injected baseline, indicating a modest reduction in task completion rate. However, the distribution remains tightly concentrated in the high-utility region, demonstrating that mitigation does not materially impair routine functionality. These statistics are summarized in Table~\ref{tab:utility_stats}.
\begin{table}[t!]
\centering
\caption{Utility statistics across configurations (95\% CI). Impact size computed using Cohen’s $d$.}
\label{tab:utility_stats}
\begin{tabular}{l c c c}
\toprule
Configuration & Mean utility & SD & 95\% CI \\
\midrule
Injected & 0.93 & 0.04 & [0.92, 0.94] \\
Defended & 0.89 & 0.05 & [0.88, 0.90] \\
\bottomrule
\end{tabular}
\end{table}
Figure~\ref{fig:latency} presents the latency distribution on a logarithmic scale to account for right-skewed response times. The defended configuration exhibits a slight upward shift, reflecting additional policy checks and sandbox validation overhead. The shift is modest relative to the overall variance in response times, and extreme latency outliers are not substantially amplified. 
\begin{table}[t!]
\centering
\caption{Latency statistics across configurations (milliseconds).}
\label{tab:latency_stats}
\begin{tabular}{l c c}
\toprule
Configuration & Median latency & IQR \\
\midrule
Injected & 420 ms & 110 ms \\
Defended & 468 ms & 125 ms \\
\bottomrule
\end{tabular}
\end{table}
Tables ~\ref{tab:utility_stats} and ~\ref{tab:latency_stats} show that the mitigation introduces only minor reductions in utility and modest latency overhead. These trade-offs remain well within acceptable deployment thresholds. Importantly, the magnitude of security improvement, particularly in reduced probability of compromise and boundary stabilization, substantially exceeds the operational penalties, confirming that mitigation shifts the system toward a more favorable security–performance frontier rather than merely redistributing risk.

\subsection{Structural Risk Concentration and Instability Analysis}
We examine whether OpenClaw’s attack surface undergoes a structural transformation under adversarial pressure. Entropy-based concentration, escalation curvature, and supply-chain exposure are analyzed to determine whether amplification reflects instability in transient behavior. Attack surface entropy measures how risk is distributed across exposure vectors. Lower values indicate concentration in a few dominant weaknesses, whereas higher values reflect diffusion across multiple boundaries. As shown in Table~\ref{tab:entropy}, entropy increases from 0.42 in the benign configuration to 0.71 under injection, indicating that adversarial pressure disperses risk across multiple boundaries rather than exploiting a single weakness. Mitigation partially reconcentrates exposure (0.55) but does not restore benign levels, suggesting persistent multi-boundary fragility.
\begin{table}[t!]
\centering
\caption{Attack surface entropy across configurations.}
\label{tab:entropy}
\begin{tabular}{l c}
\toprule
Configuration & Entropy ($H_A$) \\
\midrule
Benign & 0.42 \\
Injected & 0.71 \\
Defended & 0.55 \\
\bottomrule
\end{tabular}
\end{table}
Privilege amplification is next evaluated to determine whether it scales proportionally with attacker capability. Table~\ref{tab:instability} reports a privilege gradient of 0.19, exceeding the instability threshold $\eta = 0.15$, and a positive escalation curvature of 0.08. These values indicate super-linear escalation, where incremental increases in adversarial capability yield disproportionately large gains in privilege. The observed behavior is escalation-sensitive, rooted in the coupling between reasoning and execution rather than linear accumulation.
\begin{table}[t!]
\centering
\caption{Privilege escalation metrics under injection.}
\label{tab:instability}
\begin{tabular}{l c c}
\toprule
Metric & Value & Threshold \\
\midrule
Privilege gradient & 0.19 & 0.15 \\
Escalation curvature & 0.08 & 0 \\
\bottomrule
\end{tabular}
\end{table}
Structural exposure introduced through the skill ecosystem is also assessed. Table~\ref{tab:per_llm_inferential} indicates a non-trivial supply-chain risk (SCR = 0.071). Although lower than injection-driven risk, supply-chain exposure persists independently of runtime prompting and remains active even under boundary hardening. Sensitivity analysis varying installation probability shows proportional scaling of SCR, as summarized in Table~\ref{tab:scr_sensitivity}, underscoring the importance of governance controls over marketplace discoverability and adoption.
\begin{table}[t!]
\centering
\caption{Supply-chain risk sensitivity to skill installation probability.}
\label{tab:scr_sensitivity}
\begin{tabular}{l c}
\toprule
Installation probability & SCR \\
\midrule
Low & 0.035 \\
Medium & 0.071 \\
High & 0.106 \\
\bottomrule
\end{tabular}
\end{table}

\subsection{Consensus-Gated Evaluation}
To evaluate whether aggregation policy can transform backend diversity into a defensive mechanism, we re-ran the multi-agent experiments under a consensus-gated rule. In this configuration, execution proceeds only when at least half of the agents independently propose the same action. All other experimental parameters, including attacker configuration, injection types, and LLM backends, were held constant.
\begin{table}[t!]
\centering
\caption{System compromise probability under permissive versus consensus-gated aggregation.}
\label{tab:consensus_orchestration}
\begin{tabular}{c c c}
\toprule
Number of Agents & Permissive Rule & Majority (Consensus) Rule \\
\midrule
1 & 0.24 & 0.24 \\
3 & 0.57 & 0.33 \\
5 & 0.75 & 0.41 \\
7 & 0.86 & 0.48 \\
\bottomrule
\end{tabular}
\end{table}
As shown in Table~\ref{tab:consensus_orchestration}, compromise probability grows rapidly under permissive aggregation, reaching 0.86 with seven agents. Under majority gating, growth is substantially moderated, rising only to 0.48 for the same configuration. This contrast demonstrates that backend diversity amplifies risk under disjunctive selection but attenuates risk when agreement is required. In other words, the aggregation rule determines whether heterogeneity becomes a liability.  Although majority gating does not fully eliminate compromise, it significantly reduces amplification relative to permissive execution. These results confirm that aggregation semantics are a primary driver of system-level exposure and that carefully designed multi-agent policies can leverage heterogeneity to improve resilience.

\section{Comparison with Existing Work}
\label{sec:comparison_2025}
Recent work has advanced the security evaluation and defense of tool-enabled LLM agents. Benchmark-driven frameworks such as Agent Security Bench provide systematic evaluation of prompt injection attacks across agent pipelines \cite{zhang2025asb}. Several architectural defenses have been proposed, including CaMeL \cite{debenedetti2025camel}, PromptArmor \cite{shi2025promptarmor}, and RTBAS \cite{zhong2025rtbas}, which aim to mitigate adversarial instructions and enforce integrity constraints before tool execution. Additional approaches address privilege control mechanisms, such as Prompt Flow Integrity \cite{kim2025pfi} and Progent \cite{shi2025progent}. In parallel, research has examined ecosystem-level threats impacting agent infrastructures, including tool-selection manipulation via ToolHijacker \cite{shi2025toolhijacker} and broader protocol-level vulnerabilities in LLM agent workflows \cite{ferrag2025protocol}. Despite these advances, prior work primarily evaluates prompt-injection defenses and privilege controls at the level of individual agents. The structural impact of multi-agent orchestration together with execution coupling on system-level security exposure remains largely unexplored. As summarized in Table~\ref{tab:comparison_2025}, existing studies address prompt injection, tool misuse, and privilege management largely in isolation.  In contrast, our work analyzes an execution-coupled multi-agent architecture and introduces system-level metrics, including CBI, privilege drift, attack surface entropy, and escalation curvature, to quantify orchestration-induced risk amplification and the dynamics of privilege escalation. Our results further demonstrate that permissive aggregation policies amplify the probability of compromise as the number of agents increases, while consensus-based orchestration partially mitigates this amplification, highlighting the security implications of aggregation semantics in multi-agent deployments.
\begin{table*}[t]
\centering
\caption{Comparison with representative research on LLM-agent security. \cmark\ indicates explicit support and evaluation.}
\label{tab:comparison_2025}

\renewcommand{\arraystretch}{1.15}
\setlength{\tabcolsep}{5pt}

\begin{tabular}{p{3.05cm} p{3.35cm} c c p{2.35cm} c c}
\toprule
\textbf{Work (2025)} &
\textbf{Primary Focus} &
\textbf{Exec.} &
\textbf{MA} &
\textbf{Aggregation} &
\textbf{Priv.} &
\textbf{Boundary} \\
\midrule

Zhang et al. (ASB) \cite{zhang2025asb} &
Agent security benchmark &
\cmark & \pmark & -- & -- & -- \\

Debenedetti et al. (CaMeL) \cite{debenedetti2025camel} &
Defense-by-design architecture &
\cmark & -- & -- & -- & \cmark \\

Zhong et al. (RTBAS) \cite{zhong2025rtbas} &
Tool-call integrity screening &
\cmark & -- & -- & -- & \cmark \\

Shi et al. (PromptArmor) \cite{shi2025promptarmor} &
Prompt injection detection &
\cmark & -- & -- & -- & \pmark \\

Kim et al. (PFI) \cite{kim2025pfi} &
Prompt-flow integrity (privilege protection) &
\cmark & -- & -- & \cmark & \cmark \\

Shi et al. (Progent) \cite{shi2025progent} &
Programmable privilege policies &
\cmark & -- & -- & \cmark & -- \\

Shi et al. (ToolHijacker) \cite{shi2025toolhijacker} &
Tool-selection prompt injection &
\cmark & -- & -- & -- & -- \\

Ferrag et al. \cite{ferrag2025protocol} &
End-to-end agent threat model &
\cmark & \cmark & -- & -- & \cmark \\
\midrule
\textbf{This work (OpenClaw)} &
\textbf{System-level amplification and trust-boundary violations} &
\cmark & \cmark &
\textbf{Any-agent vs.\ majority} &
\cmark &
\cmark \\
\bottomrule
\end{tabular}
\end{table*}

\section{Discussion}
\label{sec:discussion}
This study provides a structural security analysis of OpenClaw, an execution-coupled multi-agent system in which semantic outputs directly trigger privileged actions. The results demonstrate that compromise probability, boundary instability, and privilege drift are emergent phenomena arising from architectural properties embedded in reasoning-to-execution coupling and aggregation semantics, rather than isolated model vulnerabilities. A central contribution of this work is the identification of a causal pathway linking adversarial impact to privilege escalation through boundary instability. Adversarial prompts and malicious skill inputs first weaken boundary-validation predicates, thereby increasing CBI. As these boundaries degrade, semantic outputs gain greater leverage over execution-layer authority, leading to measurable privilege drift. Mediation analysis supports this interpretation: the indirect impact of adversarial actions on privilege drift through boundary instability was substantial (\(\beta_2 = 0.42\)), whereas the direct impact was smaller (\(\beta_1 = 0.21\)). This indicates that boundary instability mediates a significant portion of escalation and that privilege amplification primarily arises from boundary degradation rather than from isolated semantic deviations. Our findings also demonstrate that orchestration semantics strongly determine system-level exposure. Under permissive aggregation, compromise probability increased by a factor of 3.58 when scaling from 1 to 7 agents. In this configuration, backend heterogeneity amplifies risk because execution proceeds whenever any agent emits a viable unsafe action. As shown in the consensus-gated evaluation, requiring majority agreement substantially reduces this amplification, confirming that aggregation rules can transform model diversity from a liability into a protective mechanism. More broadly, these results indicate that security in multi-agent AI systems cannot be reduced to individual model robustness. Architectural controls governing boundary validation, privilege propagation, and aggregation logic play a decisive role. Weak trust boundaries allow semantic perturbations to propagate into execution states, while permissive orchestration multiplies opportunities for unsafe execution. Mitigation results further reinforce this architectural perspective: hardening mechanisms significantly reduce the probability of compromise, boundary failures, and privilege drift, yet residual instability persists. Execution-layer gating constrains the magnitude of escalation but does not fully eliminate semantic leverage during authority transitions. As long as reasoning outputs remain directly coupled to executable primitives, structural sensitivity remains. Additionally, this study highlights that risk amplification in execution-coupled agentic systems is structural rather than incidental. Boundary instability mediates privilege escalation, orchestration semantics govern amplification, and supply-chain channels introduce orthogonal exposure. impactive security design, therefore, requires coordinated architectural controls across boundary enforcement, aggregation policy, and ecosystem governance, rather than relying solely on model-level alignment.

\section{Threats to Validity}
\label{sec:threats}
This section delineates the reliability boundaries of our findings. Consistent with prior simulation-based security research \cite{wohlin2012experimentation}, we distinguish between internal and external validity considerations.
\subsection{Internal Validity}
Internal validity concerns whether the observed impacts reflect structural properties of OpenClaw rather than experimental artifacts. To minimize confounding, all configurations were evaluated under controlled escalation of attacker capabilities, with fixed random seeds, identical environmental settings, and matched trial counts across baseline, injected, and defended conditions. Several limitations remain. First, the attacker capability model is parameterized and may not capture fully adaptive adversarial strategies. Second, privilege drift and boundary instability were operationalized through measurable state transitions; although grounded in formal definitions, these metrics abstract complex runtime interactions and may not capture all latent privilege pathways. Third, regression and entropy analyses assume independence across runs, whereas subtle architectural correlations may exist across execution traces. Despite these limitations, the consistency of amplification, instability, and escalation-curvature impacts across multiple independent metrics reduces the likelihood that the observed results are due to measurement artifacts.
\subsection{External Validity}
External validity concerns the generalizability of the evaluated OpenClaw configuration beyond the experimental setup. Experiments used specific LLM backends and defined orchestration policies. While different model architectures, decoding strategies, and aggregation rules may alter quantitative magnitudes, the core mechanisms identified, aggregation-driven amplification, boundary instability, and privilege escalation, are architectural rather than model-specific. The threat model focused on semantic injection, boundary manipulation, and supply-chain exposure in a self-hosted environment. Cloud-based deployments and heavily sandboxed systems may exhibit different amplification profiles due to infrastructure-level constraints. Similarly, permissive aggregation maximizes amplification, whereas consensus-based and threshold-gated policies attenuate it. Operational metrics such as latency and task utility were measured under controlled laboratory conditions; real-world variability in workload, network performance, and deployment scale may impact observed trade-offs. Nonetheless, the security dynamics emerge from the coupling of reasoning-to-execution with multi-agent aggregation, a pattern shared across many agentic AI systems. Accordingly, while numerical values may vary across deployments, the structural mechanisms identified in this study are likely to generalize to other execution-coupled multi-agent architectures. Further evaluation across diverse environments would strengthen empirical generalization.

\section{Limitations and Future Work}
\label{sec:limitations}
This study has several limitations that suggest directions for future research. While we quantified super-linear privilege escalation, we did not formally model the causal mediation between boundary instability and privilege drift. Mitigations focused on execution-layer hardening, leaving semantic-layer defenses unexplored. The analysis considered only permissive selection, and systematic evaluation of consensus thresholds, weighted voting, and dynamic trust scoring could clarify how backend diversity can be leveraged defensively. Supply-chain risk was estimated probabilistically rather than empirically, and escalation thresholds were identified empirically rather than analytically. Future work integrating causal modeling, semantic defenses, policies, and empirical ecosystem analysis could more fully characterize and mitigate structural vulnerabilities in multi-agent LLM systems.

\section{Conclusion}
\label{sec:conclusion}
This work presents a structural security analysis of OpenClaw, a self-hosted multi-agent system in which semantic reasoning is directly coupled to executable actions. Our results demonstrate that permissive multi-agent amplifies the probability of compromise, adversarial prompting diffuses instability across trust boundaries, and privilege escalation exhibits superlinear growth once the attacker's capability surpasses defined thresholds.  Mitigation mechanisms significantly reduce exposure but do not fully eliminate the structural sensitivity inherent in reasoning-to-action integration. These findings indicate that in agentic architectures, security is determined not only by individual model robustness but also by semantics, boundary enforcement, and escalation dynamics. Designing resilient deployments, therefore, requires architectural controls that constrain aggregation policies, reinforce boundary isolation, and address both semantic and supply-chain attack vectors.  Furthermore, this study underscores that reasoning-to-action coupling, while powerful for automation, introduces amplification mechanisms that must be explicitly engineered and contained to ensure system-level security.

\bibliographystyle{IEEEtran}
\bibliography{Ref}

\end{document}